\documentclass[prd,aps,nofootinbib,twocolumn,superscriptaddress,preprintnumbers,balancelastpage,longbibliography]{revtex4-1}

\usepackage{amsfonts, amssymb, amsmath}
\usepackage{graphics, graphicx}
\usepackage{booktabs, tabularx, makecell, multirow, xspace}
\usepackage[dvipsnames]{xcolor}
\usepackage[utf8]{inputenc}
\usepackage{url}
\usepackage{hyperref}
\usepackage{lipsum}

\newcommand{\LCDM}{$\Lambda$CDM\xspace}
\newcommand{\Fermi}{\emph{Fermi}\xspace}

\renewcommand{\arraystretch}{1.8}
\newcolumntype{C}[1]{>{\centering\let\newline\\\arraybackslash\hspace{0pt}}m{#1}}

\newcommand{\dd}{\mathrm{d}}
\newcommand {\be} {\begin {equation}}
\newcommand {\ee} {\end {equation}} 
\newcommand{\es}[2] {\begin{equation} \label{#1} \begin{split} #2 \end{split} \end{equation}}

\definecolor{linkcolor}{rgb}{0.7752941176470588, 0.22078431372549023, 0.2262745098039215}

\newcommand\Tstrut{\rule{0pt}{2.6ex}}         % = `top' strut
\newcommand\Bstrut{\rule[-0.9ex]{0pt}{0pt}}   % = `bottom' strut

\hypersetup{colorlinks=true
,urlcolor=linkcolor
,anchorcolor=linkcolor
,citecolor=linkcolor
,filecolor=linkcolor
,linkcolor=linkcolor
,menucolor=linkcolor
,linktocpage=true
,pdfproducer=medialab}

\begin{document}

\title{Harnessing the Population Statistics of Subhalos to \\
Search for Annihilating Dark Matter }
\author{Jean J. Somalwar}
\email{somalwar@princeton.edu}
\thanks{ORCID: \href{https://orcid.org/0000-0001-8426-5732}{0000-0001-8426-5732}}
\affiliation{Department of Physics, Princeton University, Princeton, NJ 08544}

\author{Laura J. Chang}
\email{ljchang@princeton.edu}
\thanks{ORCID: \href{https://orcid.org/0000-0001-8590-2043}{0000-0001-8590-2043}}
\affiliation{Department of Physics, Princeton University, Princeton, NJ 08544}

\author{Siddharth Mishra-Sharma}
\email{sm8383@nyu.edu}
\thanks{ORCID: \href{https://orcid.org/0000-0001-9088-7845}{0000-0001-9088-7845}}
\affiliation{Center for Cosmology and Particle Physics, Department of Physics, New York University, New York, NY 10003, USA}

\author{Mariangela Lisanti}
\email{mlisanti@princeton.edu}
\thanks{ORCID: \href{https://orcid.org/0000-0002-8495-8659}{0000-0002-8495-8659}}
\affiliation{Department of Physics, Princeton University, Princeton, NJ 08544}

\date{\today}

\begin{abstract}
 The Milky Way's dark matter halo is expected to host numerous low-mass subhalos with no detectable associated stellar component.  Such subhalos are invisible unless their dark matter annihilates to visible states such as photons.  One of the established methods for identifying candidate subhalos is to search for individual unassociated gamma-ray sources with properties consistent with the dark matter expectation.  However, robustly ruling out an astrophysical origin for any such candidate is challenging.  In this work, we present a complementary approach that harnesses information about the entire population of subhalos---such as their spatial and mass distribution in the Galaxy---to search for a signal of annihilating dark matter.  Using simulated data, we show that the collective emission from subhalos can imprint itself in a unique way on the statistics of observed photons, even when individual subhalos may be too dim to be resolved on their own.  Additionally, we demonstrate that, for the models we consider, the signal can be identified even in the face of unresolved astrophysical point-source emission of extragalactic and Galactic origin. This establishes a new search technique for subhalos that is complementary to  established methods, and that could have important ramifications for gamma-ray dark matter searches using observatories such as the \emph{Fermi} Large Area Telescope and the Cherenkov Telescope Array.
\end{abstract}

\maketitle

\section{Introduction}
\label{sec:intro}

In the Lambda Cold Dark Matter (\LCDM) framework, hierarchical clustering of matter and bottom-up structure formation predicts the clumping of dark matter across a large range of scales.  One consequence of this is that the Milky Way's dark matter (DM) halo (of total mass $\sim{10^{12}\,\mathrm{M}_{\odot}}$) is peppered with smaller ($\lesssim{10^{10}\,\mathrm{M}_{\odot}}$) ``subhalos'' in its interior.  DM subhalos within the Milky Way make ideal targets for signals of annihilating DM as they are relatively dense by definition, and hence are brighter than, \emph{e.g.}, the smooth Milky Way DM halo at the same distance.  For the case of Weakly Interacting Massive Particles (WIMPs), the photons resulting from DM annihilation have energies in the GeV range and can be searched for with gamma-ray observatories like the \Fermi Large Area Telescope~(LAT)~\cite{Atwood:2009ez}.  In this paper, we present a new approach to search for a population of subhalos in the Milky Way that takes advantage of both the bright, resolvable subhalos and those individual sources that are faint and unresolvable.

Those dark matter subhalos that have experienced star formation---known as dwarf satellite galaxies---can be identified by searches for stellar overdensities~\cite{Bechtol:2015cbp,Drlica-Wagner:2015ufc,Koposov:2015cua}, and their DM content can be inferred from, \emph{e.g}., their stellar kinematics~\cite{Kirby:2017bxa, Geha:2008zr, Martin:2007ic}. There are currently $\sim60$ confirmed and candidate dwarf galaxies identified in the Milky Way~\cite{Fermi-LAT:2016uux,Drlica-Wagner:2019vah}, and searches for annihilation signatures in these spatially localized targets have yielded some of the strongest constraints on DM annihilation to date~\cite{Ackermann:2013yva,Geringer-Sameth:2014qqa, Ackermann:2015zua,Fermi-LAT:2016uux}. Although future optical imaging surveys such as the Legacy Survey of Space and Time (LSST) on the Vera C. Rubin Observatory may discover tens to hundreds of new dwarf galaxies~\cite{Ivezic:2008fe, Drlica-Wagner:2019vah}, annihilation limits from these targets are not expected to improve dramatically~\cite{Ando:2019rvr}. Additionally, assumptions about, \emph{e.g.}, the dwarf halo shape~\cite{Geringer-Sameth:2014qqa,Sanders:2016eie,Ando:2020yyk} and stellar membership criteria used to infer the halo properties~\cite{Geringer-Sameth:2014yza,Bonnivard:2015vua} can affect the constraints obtained and the interpretation of a potential signal.

\begin{figure*}[t]
\centering
\includegraphics[width=0.45\textwidth]{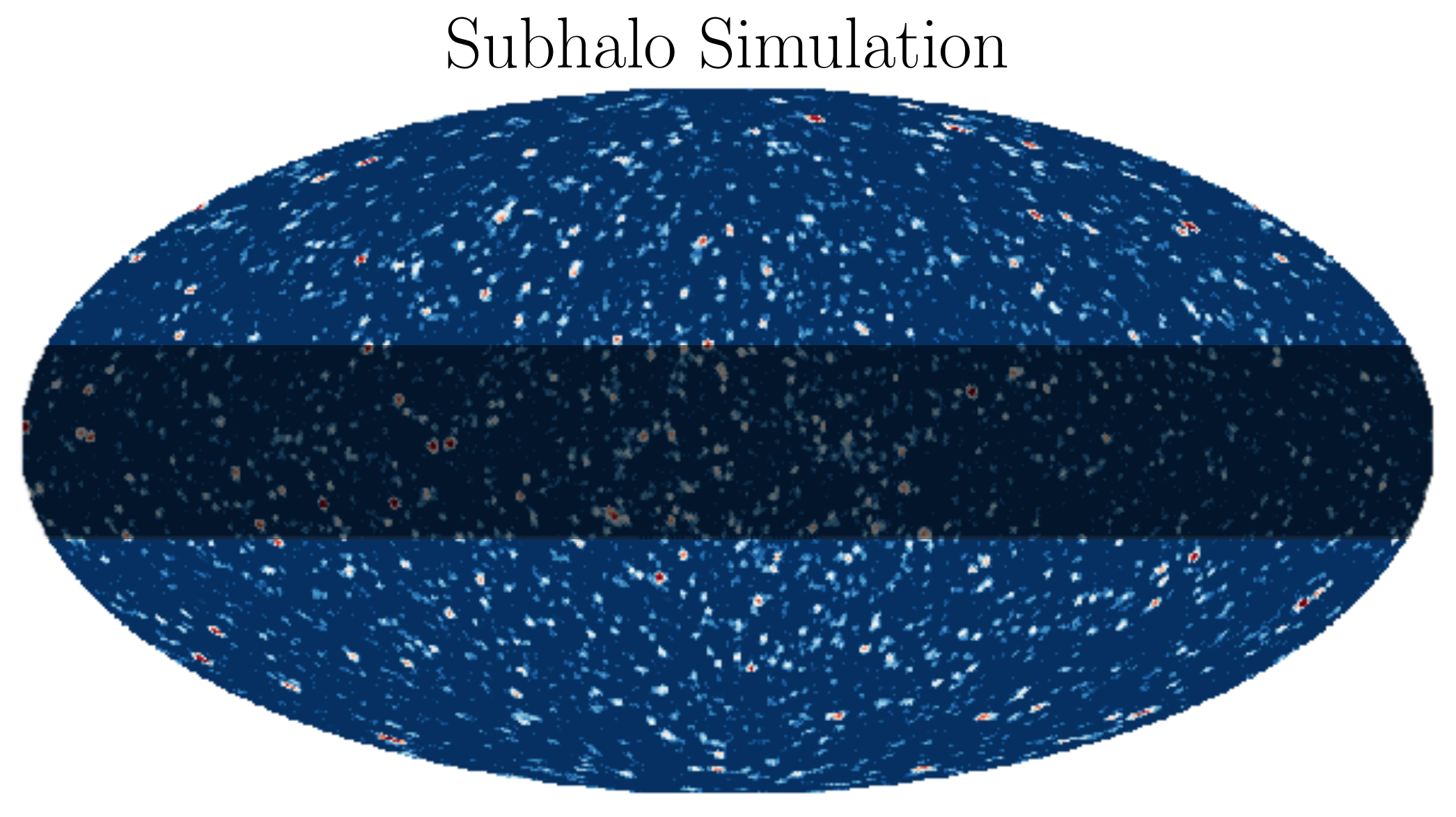}
\includegraphics[width=0.45\textwidth]{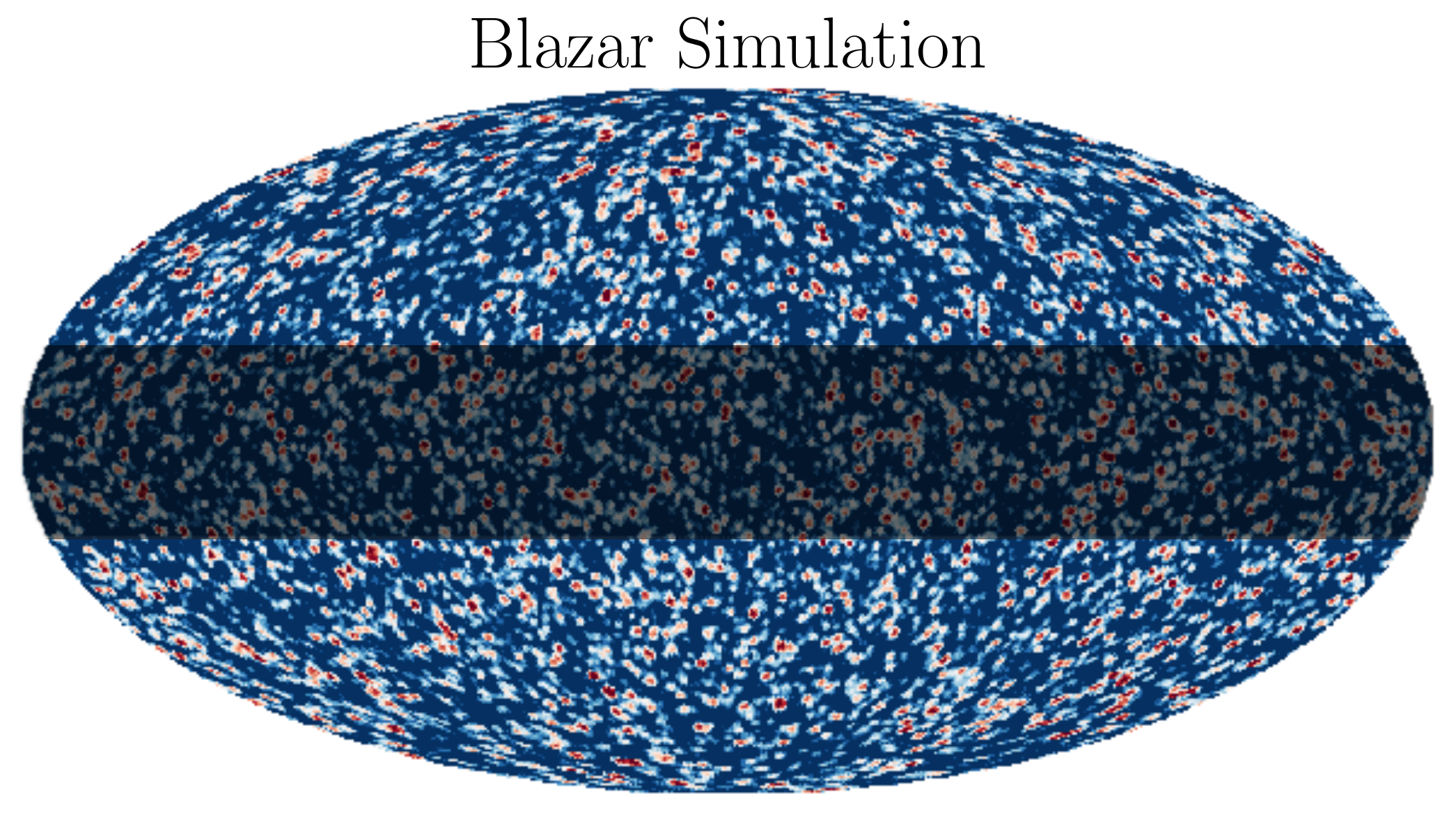}
\caption{Simulated realizations of subhalo DM annihilation (\textit{left}) and blazar (\textit{right}) count maps corresponding to our fiducial \emph{Fermi} dataset within energy range 2--20\,GeV. The dark band in each map corresponds to the region that is masked in our fiducial analysis. The maps are both smoothed with a symmetric Gaussian beam with full-width-half-maximum of $0.5^\circ$ to aid in visualization. The subhalo maps are generated from the model described in Sec.~\ref{sec:shmodels} and assume an annihilation cross section $\langle \sigma v \rangle = 10^{-24}\,{\rm cm}^3\,{\rm s}^{-1}$ and a DM mass $m_{\chi}=40$\,GeV. The blazar emission is modeled following Ref.~\cite{Ajello:2015mfa}. Both maps share a common color scale. It can be seen that blazars are both more numerous as well as brighter than the subhalos on average, with both the blazar and subhalo distributions appearing nearly isotropic over the sky.
}
\label{fig:maps}
\end{figure*}

Some subhalos, typically those that are less massive ($\lesssim 10^{7-8}\,\mathrm{M}_\odot$), can be largely devoid of baryonic activity and not detectable through searches for stellar overdensities~\cite{Brown:2014vpa, Wetzel:2015uya, Akins:2020uya}.  Much remains unknown about the expected properties of these subhalos, including the density distribution of DM in each subhalo, as well as the spatial distribution of the subhalos within the Galaxy. Although both properties have been studied using numerical simulations~\cite{Springel:2008cc,Madau:2008fr,Mollitor:2014ara,Sawala:2015cdf,Wetzel:2016wro}, large uncertainties remain, especially regarding the effects of tidal stripping during infall and that of the Galactic baryonic disk~\cite{Fitts:2018ycl,Garrison-Kimmel:2017zes,Brooks:2018ktu,Nadler:2017dxq} on the present-day shape and distribution of subhalos.

A DM annihilation signal from a subhalo could be detected as a gamma-ray source, and prospects of searching for annihilation in Galactic subhalos through such individual detections have been studied extensively~\cite{Kuhlen:2008aw,Buckley:2010vg,Belikov:2011pu,Zechlin:2011wa,2011arXiv1110.4744N,Ackermann:2012nb,Zechlin:2012by,Berlin:2013dva,Bertoni:2015mla,Schoonenberg:2016aml,Bertoni:2016hoh,Mirabal:2016huj, Hooper:2016cld,Calore:2016ogv,Hutten:2019tew, Glawion:2019fvo, Calore:2019lks,Coronado-Blazquez:2019pny,Coronado-Blazquez:2020dyl,Facchinetti:2020tcn, DiMauro:2020uos}.  These studies rely on looking for individually resolved gamma-ray sources whose emission cannot be attributed to known astrophysical activity and that have properties consistent with DM annihilation---\emph{e.g.}, energy spectrum, spatial extension, and steady, non-variable emission.  Then, the number of unassociated candidate sources can set constraints on the DM parameter space as no subhalo model should over-predict this number. In the 3FGL catalog of resolved sources, $\sim 15$--$20$ unassociated sources have been identified with properties consistent with emission from annihilating DM~\cite{Bertoni:2015mla, Schoonenberg:2016aml, Hooper:2016cld}. Recently, Ref.~\cite{Coronado-Blazquez:2019puc} identified 44 such candidate sources in the 3FGL, 2FHL, and 3FHL PS catalogs, though a follow-up spectral analysis of these sources argued that none are good viable subhalo candidates~\cite{Coronado-Blazquez:2019pny}.  However, a robust determination of their nature requires confirming that the source does not emit at other wavelengths.   

Given the importance of the WIMP paradigm, it is crucial to have a number of complementary search targets and techniques to look for DM annihilation. In particular, instead of looking for emission from individually resolved subhalos, we can try to disentangle the \emph{collective} effect of a population of dim sources (including both unresolved as well as resolved, but unassociated) to the gamma-ray sky.  The importance of considering these dim subhalos is highlighted in the left panel of Fig.~\ref{fig:maps}, which shows a simulated sky map of gamma-ray emission in the 2--20\, GeV energy range from subhalos, assuming a particular subhalo model described in Sec.~\ref{sec:shmodels}, detector characteristics described in Sec.~\ref{sec:simmaps}, and DM particles with mass $m_{\chi}=40$\,GeV annihilating to the $b\overline{b}$ final state with cross section $\langle \sigma v \rangle = 10^{-24}\,{\rm cm}^3\,{\rm s}^{-1}$.  There are very few bright, resolvable sources and the probability that any one of these should be close to the Solar position is small.  Even though the dimmer subhalos may not be resolvable individually, their annihilation signature could leave a unique imprint on the spatial and spectral distribution of observed gamma-ray photons. 

Isolating the signatures of the unresolved subhalo population depends crucially on our ability to distinguish the DM signal from that of astrophysical backgrounds of both point source (PS) and diffuse nature.  For example, the nearly isotropic emission from extragalactic blazars, which is expected to provide the dominant astrophysical contribution to the Isotropic Gamma-Ray Background (IGRB)~\cite{DiMauro:2017ing,DiMauro:2015tfa,Ajello:2015mfa,Cuoco:2012yf}, may hamper our ability to infer the presence of subhalos. A simulated realization of the expected blazar contribution to the IGRB following the best-fit population model in Ref.~\cite{Ajello:2015mfa} is shown in the right panel of Fig.~\ref{fig:maps}; these are expected to be both more numerous as well as individually brighter than annihilating DM subhalos. 

Previous studies have attempted to leverage the statistical properties of the unresolved subhalo population to search for DM annihilation. For example, the contribution of DM annihilation in Galactic subhalos to the measured flux and energy spectrum of the IGRB~\cite{Ackermann:2015tah} and its 2-point photon-count distribution (angular power spectrum)~\cite{Fornasa:2012gu,Ripken:2012db,Fornasa:2016ohl,Hutten:2018wop,Ackermann:2012uf,Ando:2013ff} have been used in combination with models of the dominant astrophysical gamma-ray contributers to constrain DM annihilation properties.

Complementary to studies using the integrated emission and angular power spectrum of DM annihilation from a population of Galactic subhalos, in this paper we present a novel strategy using 1-point photon statistics to search for the  annihilation signature. Our technique takes advantage of the information in the entire population of sources, including both those that are resolved and those that are faint and unresolved. The concept of leveraging the 1-point photon-count distribution to search for DM has previously been studied in Refs.~\cite{Dodelson:2009ih,Feyereisen:2015cea} in the context of emission from extragalactic sources and in Refs.~\cite{Lee:2008fm,Koushiappas:2010fk} with application to Galactic subhalos. We introduce a method to search for signatures of DM annihilation from a Galactic subhalo population using the Non-Poissonian Template Fitting (NPTF) framework~\cite{Malyshev:2011zi,Lee:2014mza,Lee:2015fea,Mishra-Sharma:2016gis}, which has previously been applied to characterize unresolved PSs in the Inner Galaxy~\cite{Lee:2015fea,Linden:2016rcf,Leane:2019xiy,Leane:2020pfc, Leane:2020nmi,Chang:2019ars,Buschmann:2020adf} and at high latitudes~\cite{Zechlin:2016pme,Lisanti:2016jub,Zechlin:2017uzo}. Using simulations, we show that the NPTF can constrain DM annihilation from a population of subhalos in the face of astrophysical background emission. We find that using photon statistics to look for collective emission from a subhalo population can be especially promising when a large number of individual subhalo candidates are identified in PS catalogs. This establishes a method complementary to the established ones based on characterizing individual resolved PSs as subhalo candidates, as well as those based on using the measured 0-point (overall flux) and 2-point (angular power spectrum) statistics to characterize a subhalo population. 

\begin{table*}[tb]
\footnotesize
\begin{center}
\begin{tabular}{C{4cm} | C{4cm}C{4cm}C{4cm}}
\renewcommand{\arraystretch}{1}
& \textbf{Fiducial}	 & \textbf{Concentration Variant}  & \textbf{Spatial Variant}   \Tstrut\Bstrut	\\   
\Xhline{3\arrayrulewidth}
Calibration Simulation & Aquarius~\cite{Springel:2008cc} & Aquarius~\cite{Springel:2008cc} & Phat-ELVIS \cite{Kelley:2018pdy}  \\
Spatial Distribution  & Einasto (Eq.~\eqref{eq:ein}) & Einasto (Eq.~\eqref{eq:ein}) & Sigmoid-Einasto (Eq.~\eqref{eq:SigmoidEinasto})  \\
$N_{\rm calib}$ & 300 &  300 & 90 \Tstrut\Bstrut \\ 
$c(m)$ & Molin{\'e} \emph{et al.}~\cite{Moline:2016pbm} & S{\'a}nchez-Conde \emph{et al.} \cite{Sanchez-Conde:2013yxa} & Molin{\'e} \emph{et al.}~\cite{Moline:2016pbm} \Tstrut\Bstrut \\ 
\end{tabular}
\end{center}
\caption{The three subhalo models considered in this work.  The fiducial case assumes an Einasto spatial density distribution with parameters $\alpha_\mathrm{s}$ and $r_\mathrm{s}$ taken from the Aquarius simulation~\cite{Springel:2008cc}, accounting for the effects of tidal stripping.  The concentration-mass relation, $c(m)$, is also specified as well as $N_\text{calib}$, which fixes the overall normalization to the number of subhalos in the mass range $m_{200} = 10^8$--$10^{10}$\,M$_\odot$.  We consider two variants on this benchmark case.   The first uses a distance-independent concentration relation, but keeps everything else the same.  The second variant considers a Sigmoid-Einasto spatial distribution that accounts for baryonic effects from the Milky Way disk.  In all three cases, the density distribution of an individual subhalo is modeled using an Einasto profile.} 
\label{tab:models}
\end{table*}  

This paper is organized as follows. We begin in Sec.~\ref{sec:models} by detailing how DM annihilation from Galactic substructure can imprint itself onto the gamma-ray sky and contrast this signature with that from traditional astrophysical sources. We also describe our modeling of the subhalo and astrophysical PS contributions. We then describe our analysis pipeline in Sec.~\ref{sec:methodology} and show the results of our study on simulated data in Sec.~\ref{sec:results}. In Sec.~\ref{sec:fluxreg}, we explore in greater detail the distinct features of an annihilating subhalo population that sets them apart from an astrophysical PS population. We conclude in Sec.~\ref{sec:conclusions}.  Appendix~\ref{app:appendix} includes a supplementary figure.

\section{Subhalo and Astrophysical Point-Source Models}
\label{sec:models}

We begin by describing in turn our modeling of the gamma-ray emission from DM annihilation in Galactic subhalos (Sec.~\ref{sec:shmodels}) and that from astrophysical PSs (Sec.~\ref{sec:blazmodel}). We emphasize that the particulars of the subhalo and background PS models we assume are not critical to this study---the method can be adapted to search for DM annihilation from any specified model for the population of subhalos.  We explore the sensitivity of our results to variations in subhalo modeling in Sec.~\ref{sec:results}.

\subsection{Emission from Dark Matter Annihilation \\in Subhalos}
\label{sec:shmodels}

For the simplest cases, the photon flux from annihilating DM of mass $m_\chi$ and annihilation cross-section $\langle\sigma v\rangle$ is given by
\begin{equation}
\frac{\dd\Phi}{\dd E_{\gamma}} =   J \, \times \frac{\langle\sigma v\rangle}{8 \pi m_{\chi}^{2}} \, \, \sum_j \text{Br}_{j}\, \frac{\dd N_{j}}{\dd E_{\gamma}}   \,,
\label{eq:flux}
\end{equation}
where $E_\gamma$ is the photon energy, $\text{Br}_{j}$ is the branching fraction to the $j^\text{th}$ annihilation channel, and $ \dd N_{j}/\dd E_{\gamma}$ is the photon energy distribution in the $j^\text{th}$ channel, modeled here using PPPC4DMID~\cite{Cirelli:2010xx}. Throughout this study, we consider the case of annihilation into the $b \bar{b}$ channel as a benchmark example of a continuum spectrum, and take $m_\chi = 40$\,GeV. The $J$-factor encodes the astrophysical information, and is defined as the integral along the line of sight of the squared DM density for a given target:\
\es{eq:Jfactor}{
J = \int \dd s \, \dd\Omega\,\rho^{2}_\text{DM}(s,\Omega)\,,
}
where $\rho_\text{DM}$ is the DM density, $s$ is the line-of-sight distance, and $\Omega$ is the solid angle of integration.

\begin{figure*}[t]
\centering
\includegraphics[width=0.9\textwidth]{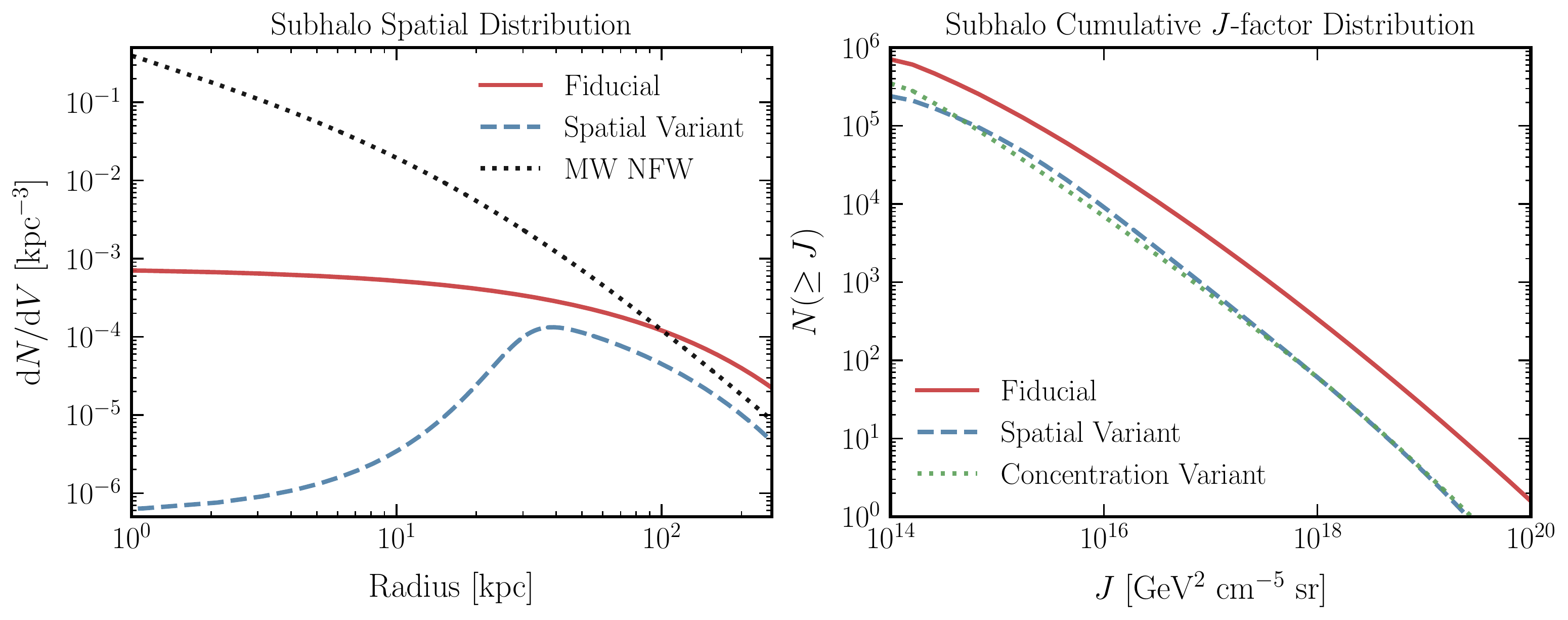}
\caption{{\it (Left)} Subhalo number density distribution for the Einasto profile fit to Aquarius (solid red line), the Phat-ELVIS Spatial Variant (dashed blue line), and the Milky Way (MW) NFW halo (dotted black line).  The number density for the Milky Way NFW halo is normalized to the same total number of subhalos as the Fiducial model.  The number of subhalos near the Solar radius ($\sim 8$\,kpc) is significantly depleted compared to the expectation from the smooth Milky Way NFW-modeled halo.  {\it (Right)}  Differential number density of subhalos with a given $J$-factor. Predictions for the Fiducial model are shown as the solid red line, for the Phat-ELVIS Spatial Variant model~\cite{Hutten:2019tew} as the dashed blue line, and for the Concentration Variant model~\cite{Sanchez-Conde:2013yxa} as the dotted green line.
}
\label{fig:Jfac}
\end{figure*}

To determine the expected gamma-ray emission from a population of subhalos, we need to model their density distribution in the Milky Way. Practically, this factorizes into modeling  their spatial distribution, their mass distribution, and the density profiles of individual subhalos. We choose as our fiducial case a benchmark model that is motivated by Refs.~\cite{Hutten:2016jko,Hutten:2019tew} and is largely based on the results of dark-matter-only Galactic simulations. We also consider two variations. First, the ``Concentration Variant'' model uses a concentration-mass relation appropriate for field halos ({\emph i.e.}, subhalos are generally less concentrated close to the Solar radius relative to our Fiducial model). Second, the ``Spatial Variant'' model includes a subhalo spatial distribution that accounts for the possibly enhanced tidal destruction of subhalos near the Galactic Center caused by the Milky Way's baryonic disk and bulge. The properties of these three models are summarized in Table~\ref{tab:models} and described in detail below.

\subsubsection{Subhalo Spatial Distribution}

A ubiquitous feature from numerical simulations is that subhalos are less centrally concentrated in the Milky Way compared to the overall smooth host halo distribution as a result of tidal effects of the Galactic potential during subhalo infall (see, \emph{e.g.}, Ref.~\cite{Han:2015pua} and references therein).  For our Fiducial model, we take the spatial distribution of subhalos from the fit by Ref.~\cite{Hutten:2019tew} to the Aq-A-1 halo of the Aquarius Project~\cite{Springel:2008cc}, a suite of DM-only simulations of Milky Way-like halos.  Specifically, the number density is modeled as an Einasto profile~\cite{Haud:1986yj}:
\es{eq:ein}{
\frac{\text{d}N}{\text{d}V}\propto\exp\left(-\frac{2}{\alpha_\mathrm{s}}\left[\left(\frac{r}{r_\mathrm{s}}\right)^{\alpha_\mathrm{s}}-1\right]\right)\,,}
with $r_\mathrm{s}=199$ kpc and $\alpha_\mathrm{s}=0.678$\,\cite{Springel:2008cc}.

As a point of comparison, our Spatial Variant model uses the spatial distribution obtained from Phat-ELVIS, a suite of DM-only simulations that were rerun with a galactic potential modeled after the Milky Way disk and bulge today~\cite{Kelley:2018pdy}. This lets us capture the effect of additional suppression in the number density of subhalos in the Inner Galaxy owing to the effects of baryons. Following Ref.~\cite{Hutten:2019tew}, we fit the number density of the surviving subhalos to the following ``sigmoid-Einasto" function:
\es{eq:SigmoidEinasto}{
\frac{\text{d}N}{\text{d}V}\propto \frac{1}{1+e^{-k(r-r_0)}}\cdot\exp\left(-\frac{2}{\alpha_\mathrm{s}}\left[\left( \frac{r}{r_\mathrm{s}}\right)^{\alpha_\mathrm{s}}-1\right]\right)\,.
}
The first term in Eq.~\eqref{eq:SigmoidEinasto} is a sigmoid function parameterized by $r_0$ and $k$, which set the midpoint and steepness of the sigmoid, respectively.  The distribution transitions to an Einasto profile at larger radii. This parameterization allows us to capture the characteristic depletion of subhalos in the inner region of the host galaxy due to the effect of the baryonic disk potential, and simultaneously match DM-only results in the outer region, where baryonic effects are negligible. We fix $\alpha_\mathrm{s}=0.68$, $r_0=29.2$~kpc, $k=0.24$, and $r_\mathrm{s}=128$\,kpc as found by Ref.~\cite{Hutten:2019tew}.

The Fiducial and Spatial Variant radial distributions are illustrated in the left panel of Fig.~\ref{fig:Jfac} as the solid red and dashed blue line, respectively.  Note that the overall normalization of the Spatial Variant number density distribution is a factor of $\sim3$ lower than that of Fiducial model because of the increased tidal disruption, as we will detail in the following section. For  comparison, we also show the expected distribution if the subhalos instead trace the overall Milky Way DM distribution---which we assume follows a Navarro-Frenk-White (NFW) profile~\cite{Navarro:1995iw} with scale radius 20\,kpc---and is normalized to match the total number of subhalos in the Fiducial model.  In both the Fiducial and Phat-ELVIS cases, the number density of subhalos is significantly depleted in the inner region of the Galaxy relative to the overall DM distribution.

\subsubsection{Subhalo Mass Function}

\LCDM simulations predict a broad, nearly scale-invariant distribution of substructure with a mass function of the form
\es{eq:massfunc}{
\frac{\text{d}N}{\text{d}m}\propto m^{-\alpha}\,,
}  with $\alpha\approx1.9$--$2$ over a large range of masses~\cite{Springel:2008cc,Moline:2016pbm}. We model the mass distribution of subhalos according to Eq.~\eqref{eq:massfunc} with $\alpha=1.9$ in our Fiducial model, normalizing the overall abundance of substructure by requiring $N_\text{calib}=300$ subhalos in the mass range $m_{200} = 10^8$--$10^{10}\,\text{M}_\odot$,\footnote{Note that we use the mass definition $m_{200}$, corresponding to the evolved subhalo mass, throughout this paper.} following Ref.~\cite{Hutten:2016jko,Springel:2008cc}. When the spatial distribution of subhalos is modeled following the Phat-ELVIS simulations (Spatial Variant model), we keep the form of the mass function the same but instead normalize to $N_\text{calib}=90$ subhalos with tidal masses in the same range, following Ref.~\cite{Hutten:2019tew}.
We include subhalos with masses in the range $m_{200}=10^{4}$--$10^{10}\,{\rm M}_{\odot}$. The upper mass range is motivated by the maximum mass of satellite galaxies consistent with simulations~\cite{Despali:2016meh,Hiroshima:2018kfv}. Subhalos with masses $<10^4\,\mathrm{M}_{\odot}$ contribute $\sim 3\%$ of the total flux in the form of smooth background emission in our Fiducial model, and hence are negligible for our purposes. These configurations result in $\approx 10^{6}$ subhalos in the fiducial case, and $\approx 4 \times 10^5$ subhalos in the Spatial Variant case. 

\subsubsection{Subhalo Density Profile}

The density profile of an individual subhalo is modeled using an Einasto profile (Eq.~\eqref{eq:ein}), 
which is fully specified by a virial mass $m_{200}$, defined as the mass contained within a virial radius $r_{200}$, which is the radius within which the mean density is 200 times the critical density of the Universe, and the virial concentration parameter $c_{200}\equiv r_{200}/r_\mathrm{s}$ relating the virial and scale radii (see, \emph{e.g.}, Ref.~\cite{Lisanti:2017qoz} for further details). 

In general, a halo's concentration strongly correlates with its mass, and semi-analytic as well as simulation-based approaches have been used to quantify this relation (\emph{e.g.}, Refs.~\cite{Sanchez-Conde:2013yxa,Diemer:2014gba,Correa:2015dva,Bartels:2015uba}). We use the concentration-mass relation from Ref.~\cite{Moline:2016pbm} in our Fiducial model, which takes into account the different environment of subhalos compared to field halos and models the effect of tidal disruption on subhalo concentrations. In particular, this results in a concentration-mass relation that is dependent on the Galactocentric distance of the subhalo, with subhalos closer to the Galactic Center being more concentrated due to tidal effects. As a point of reference, we also consider the field halo concentration-mass relation from Ref.~\cite{Sanchez-Conde:2013yxa}, which does not model tidal effects and results in a Galactocentric distance-independent concentration-mass relation, in our Concentration Variant model.
Note that we do not consider boost due to sub-substructure, which could further enhance the annihilation signal compared to the scenarios presented here~\cite{Hutten:2016jko,Ando:2019xlm,Ishiyama:2019hmh}.

The right panel of Fig.~\ref{fig:Jfac} shows the cumulative distribution of $J$-factors for the three subhalo models considered. Compared to the Galactocentric distance-independent concentration-mass relation of Ref.~\cite{Sanchez-Conde:2013yxa}, the Fiducial model gives a higher number of subhalos across the entire range of $J$-factors due to typically enhanced subhalo concentrations closer to the Solar position. The Spatial Variant model based on the Phat-ELVIS spatial distribution depletes the overall number of objects across the entire $J$-factor range due to interaction of subhalos with the baryonic potential.  The Concentration Variant model also has a reduced number of subhalos across all $J$-factors due to the relatively lower concentrations of the subhalos near the Solar radius.

The source-count distribution---which quantifies the differential number density of point sources as a function of observed flux---in the 2--20\,GeV energy range from DM annihilation for our Fiducal model is shown in the left panel of Fig.~\ref{fig:SCD_all} for an assumed DM mass $m_{\chi}=40$\,GeV, $b\overline{b}$ final state, and annihilation cross sections $\langle\sigma v\rangle=[10^{-25},10^{-24},10^{-23}]$\,cm$^3$\,s$^{-1}$. For comparison, the right panel of Fig.~\ref{fig:SCD_all} shows the source-count distributions for the variations on the subhalo model considered, fixing $\langle\sigma v\rangle=10^{-24}$\,cm$^3$\,s$^{-1}$.

\subsection{Emission from Astrophysical Point Sources}
\label{sec:blazmodel}
\begin{figure*}[t]
\centering
\includegraphics[width=0.9\textwidth]{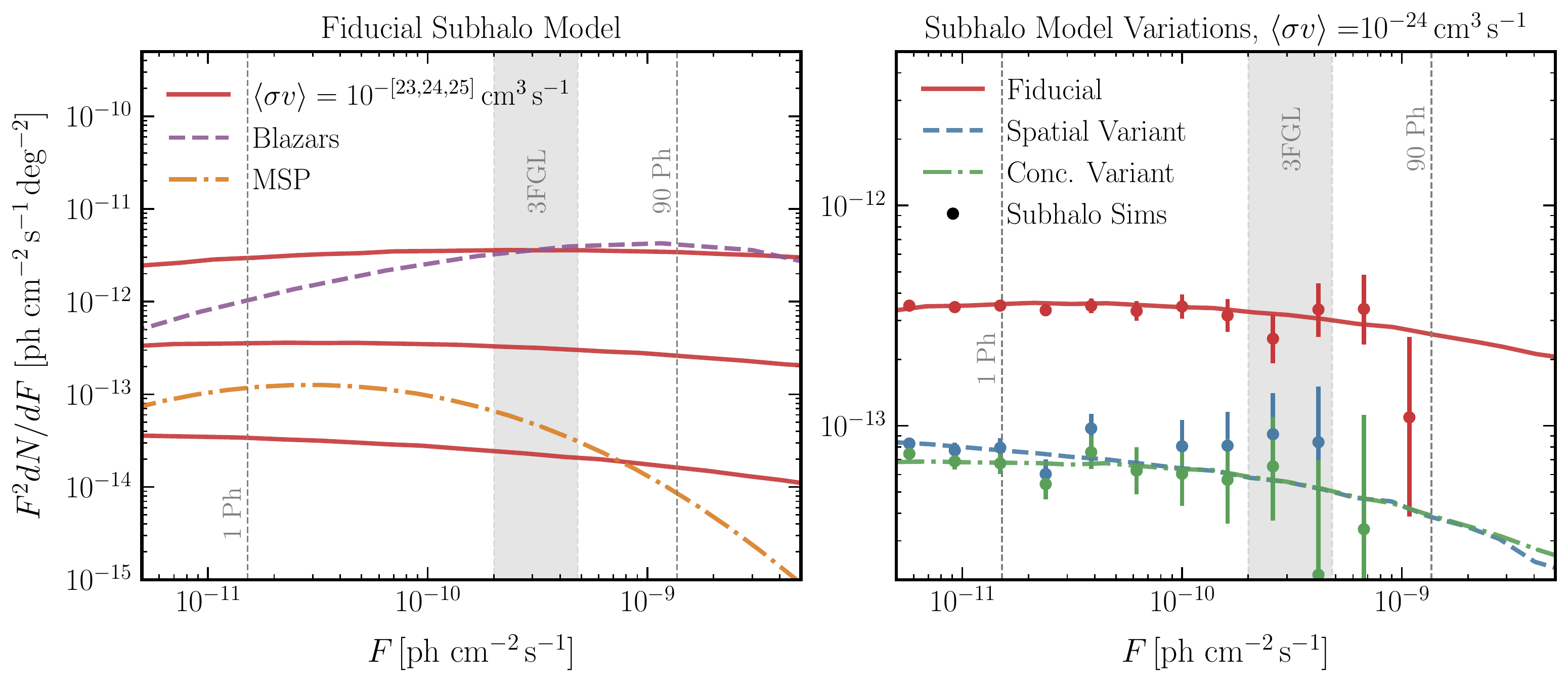}
\caption{ \emph{(Left)}  Source-count distributions in the  2--20~GeV energy range for the Fiducial subhalo model, shown for cross sections $\langle\sigma v\rangle=[10^{-25},10^{-24},10^{-23}]$\,cm$^3$\,s$^{-1}$ in solid red. DM mass $m_\chi = 40$\,GeV is assumed. The assumed theoretical blazar and Galactic pulsar distributions are shown in dashed purple and dot-dashed orange, respectively. \emph{(Right)} Source-count distributions for the Fiducial subhalo model (solid red) as well as the Spatial (dashed blue) and Concentration (dot-dashed green) variations considered, assuming $\langle\sigma v\rangle=10^{-24}$\,cm$^3$\,s$^{-1}$.  Single realizations of subhalo number counts for each model are shown as the data points along with statistical errors.  In both panels, the grey band shows the approximate detection threshold of the 3FGL PS catalog obtained using the \emph{Fermi} source detection efficiency. The 1-photon line is the approximate threshold to which the NPTF is sensitive; the 90-photon line demarcates the approximate threshold above which all sources are assumed to be associated.
}
\label{fig:SCD_all}
\end{figure*}
The dominant background when searching for signatures of DM annihilation at higher latitudes is expected to be emission from unresolved astrophysical sources of both extragalactic and Galactic origin. Inferring the astrophysical origin of this unresolved gamma-ray background (UGRB) is an active ongoing area of research, and modeling this in full generality is beyond the scope of this work.  We will instead use a simplified background PS model in this study, which we motivate here.

The extragalactic component of the UGRB---the IGRB---has been assumed to be comprised of different astrophysical source classes, in particular blazars, star-forming galaxies, and radio galaxies~\cite{DiMauro:2014wha,DiMauro:2015tfa,DiMauro:2013xta,Ajello:2011zi,DiMauro:2013zfa,Tamborra:2014xia,Hooper:2016gjy,Linden:2016fdd,Ajello:2015mfa}. These interpretations necessarily involve extrapolating the properties of resolved sources below the \emph{Fermi} resolution threshold and, in the case of star-forming and radio galaxies, assumptions about the correlation of gamma-ray luminosities with infrared and radio emission, respectively. Even with the large uncertainties that follow from these assumptions, there is strong indication that the UGRB at high latitudes ($\gtrsim 20^\circ$) and above $\sim\text{GeV}$ energies is predominantly of extragalactic origin and significantly composed of emission from blazar-like sources~\cite{DiMauro:2015tfa,Ajello:2015mfa}. This has been confirmed by data-driven methods based on photon statistics~\cite{Lisanti:2016jub,Manconi:2019ynl,Zechlin:2016pme} and using corrections to the \emph{Fermi} efficiency function at low fluxes to infer the intrinsic properties of the blazar-like population from the observed source-count distribution~\cite{Marcotulli:2020fpm,DiMauro:2017ing,TheFermi-LAT:2015ykq}. These results are also compatible with interpretations of the measured \emph{Fermi} angular power spectrum~\cite{Ackermann:2018wlo,Ando:2017alx,Cuoco:2012yf,Harding:2012gk}. Taken together, these studies point to the fact that blazars make up at least $\sim 50$--$80\%$ (depending on the energy range) of the integrated extragalactic gamma-ray background, and that while other source classes (\emph{e.g.}, star-forming galaxies) may constitute a non-trivial part of the IGRB, the majority of the emission just below PS resolution threshold---which is most relevant to this study---is of blazar-like origin.

We therefore model extragalactic PS emission as originating from blazars following Ref.~\cite{Ajello:2015mfa}, which uses the first \emph{Fermi} AGN catalog~\cite{Abdo:2010ge} and considers the blazar sub-classes of BL Lacs and Flat-Spectrum Radio Quasars (FSRQs) together in order to improve statistics, fitting for the modeled luminosity function and energy spectrum of unresolved blazars. We refer the reader to Ref.~\cite{Lisanti:2016jub} for further details on inferring the blazar source-count distribution from these ingredients. The source-count distribution of blazars for our assumed model is shown in purple in the left panel of Fig.~\ref{fig:SCD_all}. It can be seen that the number density of blazars is expected to be several orders of magnitude larger than that of DM annihilation from subhalos for annihilation cross sections $\langle\sigma v\rangle \lesssim 10^{-24}$\,cm$^3$\,s$^{-1}$ for our Fiducial DM model. Therefore, given the nearly-isotropic spatial distribution of subhalos (see Fig.~\ref{fig:maps}), the efficacy of our method will rely crucially on the ability of the analysis technique to distinguish between the unique source-count distributions of the two populations. We treat the emission from non-blazar sources (\emph{e.g.}, star-forming galaxies) as purely smooth and isotropic, consistent with the results of Ref.~\cite{Lisanti:2016jub}.

Galactic PSs may also constitute an important background for subhalo population searches. Although expected to be subdominant to extragalactic PSs in terms of overall emission, the (non-isotropic) spatial distribution of Galactic pulsars may more closely follow that of subhalos, introducing degeneracies with the signal description. This contribution may be modeled in a similar fashion to that of blazars; for simplicity, we exclude it from our fiducial set-up and study the effect of including an unmodeled pulsar population in Sec.~\ref{sec:results}. Pulsars are simulated using the best-fit model from Ref.~\cite{Bartels:2018xom}, and the inferred source-count distribution in our ROI is shown as the orange line in the left panel of Fig.~\ref{fig:SCD_all}. The pulsar source-count distribution is orders of magnitude lower than that of the blazars, but may still affect sensitivity to a subhalo signal.

\section{Statistical Methods}  
\label{sec:methodology}

We use simulated data to characterize the ability of the NPTF procedure to set constraints on DM annihilation from a population of unresolved Galactic subhalos.  The method is complementary to other DM search techniques, including those utilizing individual resolved subhalo candidates~\cite{Kuhlen:2008aw,Buckley:2010vg,Belikov:2011pu,Zechlin:2011wa,2011arXiv1110.4744N,Ackermann:2012nb,Zechlin:2012by,Berlin:2013dva,Bertoni:2015mla,Schoonenberg:2016aml,Bertoni:2016hoh,Mirabal:2016huj, Hooper:2016cld,Calore:2016ogv,Hutten:2019tew, Glawion:2019fvo, Calore:2019lks,Coronado-Blazquez:2019pny,Coronado-Blazquez:2020dyl,Facchinetti:2020tcn, DiMauro:2020uos}, and we seek to compare the effectiveness of the two distinct approaches.  In this section, we review how the simulated maps are made, the NPTF framework, and the likelihood procedure. We conclude by briefly discussing the resolved subhalo candidate constraints which we will use as a benchmark to assess our analysis sensitivity.

\subsection{NPTF Procedure}
\label{sec:NPTF}

In the NPTF framework, we assume that the data can be described by a set of different gamma-ray components. Each component is specified by a ``template" that traces its spatial morphology. We include templates that model smooth emission or resolved PSs---both described by Poissonian statistics---as well as templates that trace populations of unresolved PSs, which are described by non-Poissonian statistics.

We spatially bin the data map $d$, such that it consists of the photon counts $n_p$ in each pixel $p$. Then, for a given model with free parameters $\boldsymbol{\theta}$, the likelihood function is given by
\es{eq:likelihood}{
\mathcal L(d \, | \, \boldsymbol{\theta}) = \prod_p \, p_{n_p}^{(p)}(\boldsymbol{\theta})\,,}
where $p_{n_p}^{(p)}(\boldsymbol{\theta})$ is the probability of drawing $n_p$ photons in pixel $p$ for the assumed model. In the Poissonian case, the templates (spatially binned in the same way as the data) simply predict the mean expected counts $\mu_p(\boldsymbol{\theta})$ in pixel $p$:
\es{eq:mu}{
\mu_p(\boldsymbol{\theta})=\sum_{l} \mu_{p,l}(\boldsymbol{\theta})\,,}
where $l$ is the index over templates. Then, $p_{n_p}^{(p)}(\boldsymbol{\theta})$ is the Poisson probability of observing $n_p$ photons given the expected counts, $p_{n_p}^{(p)}(\boldsymbol{\theta})= \mathrm{Pois}\left(n_p|\mu_p(\boldsymbol{\theta})\right)$.

When modeling unresolved PSs, however, $p_{n_p}^{(p)}(\boldsymbol{{\theta}})$ is not Poissonian and the resulting equations are more complicated. We present the relevant expressions for the main inputs into the NPTF likelihood here, but refer the reader to Refs.~\cite{Mishra-Sharma:2016gis,Chang:2019ars,Lee:2014mza,Lee:2015fea, Linden:2016rcf} for a more detailed discussion of the NPTF formalism and implementation. 

In addition to spatial morphology, an essential input when modeling a population of unresolved PSs is their flux distribution. We consider two methods for modeling the source-count distribution: \emph{(i)} fixing it to a numerically-specified function, such as one of the theoretical source counts shown in the left panel of Fig.~\ref{fig:SCD_all}, and \emph{(ii)} parametrically fitting for it using a multiply-broken power law. For the latter, we employ a triply-broken power law to model the differential source-count distribution of isotropically-distributed source, 
with $S_{\mathrm{b},\{1,2,3\}}$ parameterizing the locations of the breaks, $n_{\{1,2,3,4\}}$ parameterizing the power-law indices, and $A^\text{PS}$ the overall normalization (see Ref.~\cite{Mishra-Sharma:2016gis} for further details).  For computational ease, we construct the simulated data maps and perform the analysis assuming a uniform exposure map, setting the per-pixel exposure to the mean \emph{Fermi}-LAT exposure for the relevant dataset.  In this case, there is a simple relation between the photon count $S$ and the observed flux $F$, which is given by $S = \langle\epsilon\rangle F$, where $\langle\epsilon\rangle\simeq6.59\times 10^{10}$\,cm$^{2}$\,s is the mean exposure per pixel for the dataset used.  To account for a more realistic, non-uniform exposure, the source-count distribution $\dd N/\dd S$ should be written in terms of flux $\dd N/\dd F$, with the translation to counts occurring on a pixel-by-pixel basis or by subdividing the analysis region into separate uniform-exposure sections.  This procedure is described in Ref.~\cite{Mishra-Sharma:2016gis}.  The fact that we use a simple exposure map here does not affect the generality of the conclusions. We account for the non-trivial point-spread function (PSF) of the LAT instrument, modeling it as a linear combination of King functions using the parameterization at 2\,GeV provided by the \Fermi collaboration\footnote{\url{https://fermi.gsfc.nasa.gov/ssc/data/analysis/documentation/Cicerone/Cicerone_LAT_IRFs/IRF_PSF.html}} (see Ref.~\cite{Mishra-Sharma:2016gis} for further details on the implementation).

We assess the sensitivity of our methods using a frequentist profile likelihood-based framework. For a given subhalo model, we build up a likelihood profile for the DM annihilation cross section $\langle \sigma v\rangle$ by fixing the subhalo source count to the numerically computed function at each cross section while marginalizing over parameters associated with the background model at each value of $\langle \sigma v\rangle$. We consider two different ways of treating the astrophysical PS backgrounds in order to bracket extreme scenarios related to blazar epistemology: \emph{(i)} fixing the blazar source count to its known, true distribution, and \emph{(ii)} marginalizing over the blazar contribution by modeling it with a triply-broken power law. These correspond to the cases of assuming a given background PS model \emph{a priori} and remaining agnostic to any assumptions about the nature of astrophysical PS populations, respectively. In practice, we expect to have some prior knowledge about the contribution of blazars to the IGRB from a combination of observations and theoretical modeling~\cite{Marcotulli:2020fpm,DiMauro:2017ing,Manconi:2019ynl,TheFermi-LAT:2015ykq}.

The test-statistic (TS) profile over the DM annihilation cross section is defined as
\begin{equation}
\label{eq:TS}
\mathrm{TS} \equiv 2\left[\log \mathcal{L}\left(d \mid \langle\sigma v\rangle\right)-\log \mathcal{L}\left(d \mid {\langle\sigma v\rangle} = 0\right)\right]\,,
\end{equation}
where we have left the dependence on other DM model parameters (annihilation channel and particle mass) implicit. The marginal likelihood at each value of the cross section $\langle \sigma v \rangle$ considered is obtained by maximizing the likelihood with respect to the nuisance parameters $\boldsymbol\theta$:
\begin{equation}
\mathcal{L}\left(d \mid \langle\sigma v\rangle\right)=\max _{\left\{\boldsymbol{\theta}\right\}} \mathcal{L}\left(d \mid \langle\sigma v\rangle, \boldsymbol{\theta}\right) \, ,
\end{equation}
where $\boldsymbol{\theta}$ are the parameters describing the astrophysical background contributions, both Poissonian and non-Poissonian, being marginalized over. When the blazar source-count distribution is assumed to be known and fixed, these are simply the normalizations of the Poissonian templates described in the following section: $\boldsymbol{\theta} =  \{A_\mathrm{dif}, A_\mathrm{iso}, A_\mathrm{3FGL}, A_\mathrm{bub}\}$; when we are agnostic to the properties of unresolved, extragalactic PSs, they in turn include parameters characterizing an isotropic non-Poissonian PS component: $\boldsymbol{\theta} =  \left\{A_\mathrm{dif}, A_\mathrm{iso}, A_\mathrm{3FGL}, A_\mathrm{bub}, A_\mathrm{PS}^\mathrm{iso}, n_{\{1,2,3,4\}}^\mathrm{iso}, S_{\mathrm{b},\{1,2,3\}}^\mathrm{iso} \right\}$. After constructing the TS profile, the cross section limit may be obtained by thresholding at $\mathrm{TS} = -2.71$, corresponding to a 95\% confidence interval for a one-sided $\chi^2$ distribution. We use the public code \texttt{NPTFit}~\cite{Mishra-Sharma:2016gis}, modified to accept numerically-specified source-count distributions, to compute the NPTF likelihood.

We emphasize that our analysis only uses photons from a single energy bin (2--20~GeV), and therefore does not utilize information regarding the energy spectrum of the subhalo and blazar emission.  Additionally, the NPTF procedure treats each subhalo as a PS and does not account for any spatial extension of the source, which could be associated with a DM origin.  While this is sufficient for our proof-of-principle study, one can potentially extend the NPTF method to include energy and spatial extension information, thereby providing additional distinguishing handles.

\subsection{Simulated Maps}
\label{sec:simmaps}

We impose a fiducial latitude cut of ${\lvert}b{\rvert}>20^\circ$ in defining our region of interest (ROI).  This ROI avoids the Galaxy's mid-plane and inner central region, where cosmic-ray emission is particularly bright, while still spanning a large enough area to capture differences between the subhalo and blazar distributions. As we will discuss, we also test applying a latitude cut of ${\lvert}b{\rvert}>30^\circ$ to further mitigate  uncertainties in the emission near the Galactic plane. We use the datasets and templates from Ref.~\cite{rodd_nicholas_safdi_siddharth_2016} (packaged with Ref.~\cite{Mishra-Sharma:2016gis}) to create the simulated maps. The data and templates used correspond to 413 weeks of \emph{Fermi}-LAT Pass 8 data collected between August 4, 2008 and July 7, 2016. The top quarter of photons in the energy range 2--20~GeV by quality of PSF reconstruction (corresponding to PSF3 event type) in the event class \texttt{ULTRACLEANVETO} are used. The recommended quality cuts are applied, corresponding to zenith angle less than 90$^\circ$, \texttt{LAT\_CONFIG} = 1, and \texttt{DATA\_QUAL} $> 0.1$.\footnote{\url{https://fermi.gsfc.nasa.gov/ssc/data/analysis/documentation/Cicerone/Cicerone_Data_Exploration/Data_preparation.html}} The maps are spatially binned using HEALPix~\cite{Gorski:2004by} with \texttt{nside} = 128. Note that here we only use the real \Fermi-LAT counts data to determine the appropriate normalizations for the Poissonian astrophysical background templates.

The simulated data maps are a combination of smooth (\emph{i.e.}, Poissonian) and PS contributions. Each PS population is completely specified by its spatial and source-count distribution.  The details of the subhalo and astrophysical PS models that we consider are summarized in Sec.~\ref{sec:models}. Photon counts from a generated PS population are put down on a map according to the same \emph{Fermi} PSF described in the previous section using the algorithm implemented in the code package \texttt{NPTFit-Sim}~\cite{NPTFit-Sim}.

The points in the right panel of Fig.~\ref{fig:SCD_all} show the subhalo source counts from a single Monte Carlo realization of the simulated map, with statistical errors included.  As desired, the simulated source counts follow the expected theory distributions.  The gray band in Fig.~\ref{fig:SCD_all} denotes the approximate resolution threshold for the 3FGL sources for the dataset considered.  Sources that fall below this threshold are unresolved by standard PS identification methods.  Sources that fall above the threshold are resolved, but their identity may still be unknown.  These unassociated sources may be blazars or other known astrophysical sources that have not been detected in other wavebands, or they may be novel sources like subhalos.

In addition to the PS emission from subhalos and blazars, we also account for Poissonian astrophysical emission in the simulated maps.  These contributions include: \emph{(i)}~the Galactic diffuse foreground emission, described by \texttt{Model~A} from Ref.~\cite{Ackermann:2014usa} and found in that reference to be formally the best fit of the considered models to \emph{Fermi} data at higher latitudes $|b|>20^\circ$, \emph{(ii)}~isotropic emission, \emph{(iii)}~resolved PSs from the 3FGL catalog, and \emph{(iv)}~emission from the \Fermi bubbles~\cite{Su:2010qj}. The latter three templates are obtained from Ref.~\cite{rodd_nicholas_safdi_siddharth_2016} and the normalization of each template is set to the best-fit coefficient from a Poissonian regression of the templates to the \emph{Fermi} data in the analysis ROI. To get the normalization of the Poissonian isotropic template, we subtract the modeled blazar contribution from the best-fit isotropic Poissonian emission. This ensures that the total IGRB flux is consistent with observations. The final maps are obtained by combining a Poisson-fluctuated realization of the combined astrophysical templates with the subhalo and blazar PS maps. We use the provided PS mask~\cite{rodd_nicholas_safdi_siddharth_2016} to mask resolved 3FGL PSs at 95\% PSF containment.   

For our proof-of-principle study, we only consider one model for the Galactic diffuse emission.  However, the issue of mismodeling of this foreground emission must be contended with in application of the method to real data. Although expected to be less severe at the higher latitudes relevant to this analysis~\cite{Lisanti:2016jub} in comparison to the Inner Galaxy~\cite{Leane:2019xiy,Lee:2015fea}, a range of foreground models and/or data-driven techniques to mitigate mismodeling~\cite{Buschmann:2020adf} should be considered in a robust analysis on data.

When generating and analyzing the simulated maps, we only include sources that emit up to $90$\,photons, in expectation.  This is motivated by the fact that the brightest unassociated subhalo-candidate source in the 3FGL catalog emits about 90 photons in the data set considered; any brighter PSs (in our case, blazars) would have been resolved and associated in the energy range of study, and hence including them would only increase the level of background contamination. 

As an alternative, we consider using the approximate 3FGL source detection efficiency,\footnote{\url{https://fermi.gsfc.nasa.gov/ssc/data/access/lat/4yr_catalog/}} to exclude brighter, resolved sources (both astrophysical and subhalos) from the analysis. The flux detection efficiency is built up from the spatial efficiency map by finely binning the pixel-wise efficiency as a function of photon flux within our ROI. Applying this threshold corresponds to an analysis where all resolved sources, associated as well as unassociated, are masked. This scenario excludes information about resolvable subhalos; emission from bright, resolved astrophysical background sources is correspondingly reduced. We note that this detection efficiency was derived by extrapolating the spectral properties of resolved sources~\cite{Acero:2015hja}, and is purely demonstrative for the purposes of our study. Indeed, the 3FGL source detection efficiency map is computed assuming the full 1--100 GeV energy range, whereas we apply it to sources simulated assuming an energy range of 2--20 GeV.

Note that we do not consider here the DM annihilation signal from the smooth Milky Way host halo.  In the presence of a true DM signal, the DM particles in the smooth Milky Way halo will annihilate with each other to produce gamma rays. We call this the ``smooth'' component. DM particles in the smooth Milky Way halo will also interact with the subhalo DM particles to produce a ``cross-product'' component, thereby boosting the subhalo signal.  The expected per-pixel flux from the smooth component is expected to be of the same order of magnitude as that from standard astrophysical backgrounds at the lowest latitudes in our ROI ($|b|\sim 20^\circ$) for the cross sections considered here. While this low-latitude contribution motivates the use of the smooth halo as a target for annihilation searches~\cite{Chang:2018bpt,Huang:2015rlu,Ackermann:2012rg}, the cumulative effect within our ROI is sub-dominant.  Additionally, while accounting for the cross-product contribution to the DM flux may further improve sensitivity, we have checked that the effect at the higher latitudes we consider is not significant.

\subsection{Sensitivity Using Resolved Sources} \label{sec:reslimit}

An important benchmark when evaluating the effectiveness of the analysis method presented in this work---which focuses on picking up the aggregate signal from a population of resolved and unresolved PSs---is the corresponding limit derived from an established method in the literature, such as searches for individual, significantly resolved, candidate subhalos~\cite{Kuhlen:2008aw,Buckley:2010vg,Belikov:2011pu,Zechlin:2011wa,2011arXiv1110.4744N,Ackermann:2012nb,Zechlin:2012by,Berlin:2013dva,Bertoni:2015mla,Schoonenberg:2016aml,Bertoni:2016hoh,Mirabal:2016huj, Hooper:2016cld,Calore:2016ogv,Hutten:2019tew, Glawion:2019fvo, Calore:2019lks,Coronado-Blazquez:2019pny,Coronado-Blazquez:2020dyl,Facchinetti:2020tcn, DiMauro:2020uos}. While we will use methods based on individual detections as a benchmark for comparison, we emphasize the complementarity of the 1-photon approach to methods that rely on the  total flux, energy spectrum, and angular power spectrum of the IGRB.

For consistent comparison with our projected sensitivities, we compute the sensitivity of using individual detections, for a given subhalo model, using a simplified Poissonian framework. For a given source detection efficiency $t(S)$, we can calculate the total expected number of subhalos corresponding to a given signal cross section $\langle \sigma v \rangle$,
\begin{equation}
    N_\mathrm{exp}\left(\langle \sigma v \rangle\right) = \int_{0}^{\infty}\dd S\,t(S)\,\frac{\dd N}{\dd S}\left(S, \langle \sigma v \rangle\right) \, .
\end{equation}
Then, for a given number $N_{\rm cand}$ of candidate subhalos, the likelihood is simply Poissonian in the expected number of sources, $\mathcal L = \mathrm{Pois}\left(N_{\rm cand} \mid N_\mathrm{exp}(\langle \sigma v \rangle)\right)$. A test statistic analogous to Eq.~\eqref{eq:TS} can then be constructed and used to set projected constraint at the 95\% confidence level.

We assume the resolved candidate subhalos are observed in a larger ROI with ${\mid}b{\mid}>2^\circ$, since bright resolved sources can be detected closer to the Galactic plane. We follow the same procedure as that used in Sec.~\ref{sec:simmaps} to calculate the \Fermi-LAT detection efficiency in this ROI. Analyses of unassociated sources in the 3FGL catalog typically find roughly $\sim15$--$20$ subhalo candidates~\cite{Schoonenberg:2016aml, Calore:2016ogv,Hooper:2016cld,Coronado-Blazquez:2019puc}; we therefore show projected constraints assuming $N_{\rm cand}=1$ and $N_{\rm cand}=20$ in order to bracket the typical sensitivity of established methods. We will show our derived limits---which are consistent with previously published results that assume similar subhalo models~\cite{Calore:2019lks}---in Sec.~\ref{sec:results}.

\begin{figure*}[t]
\centering
\includegraphics[width=0.85\textwidth]{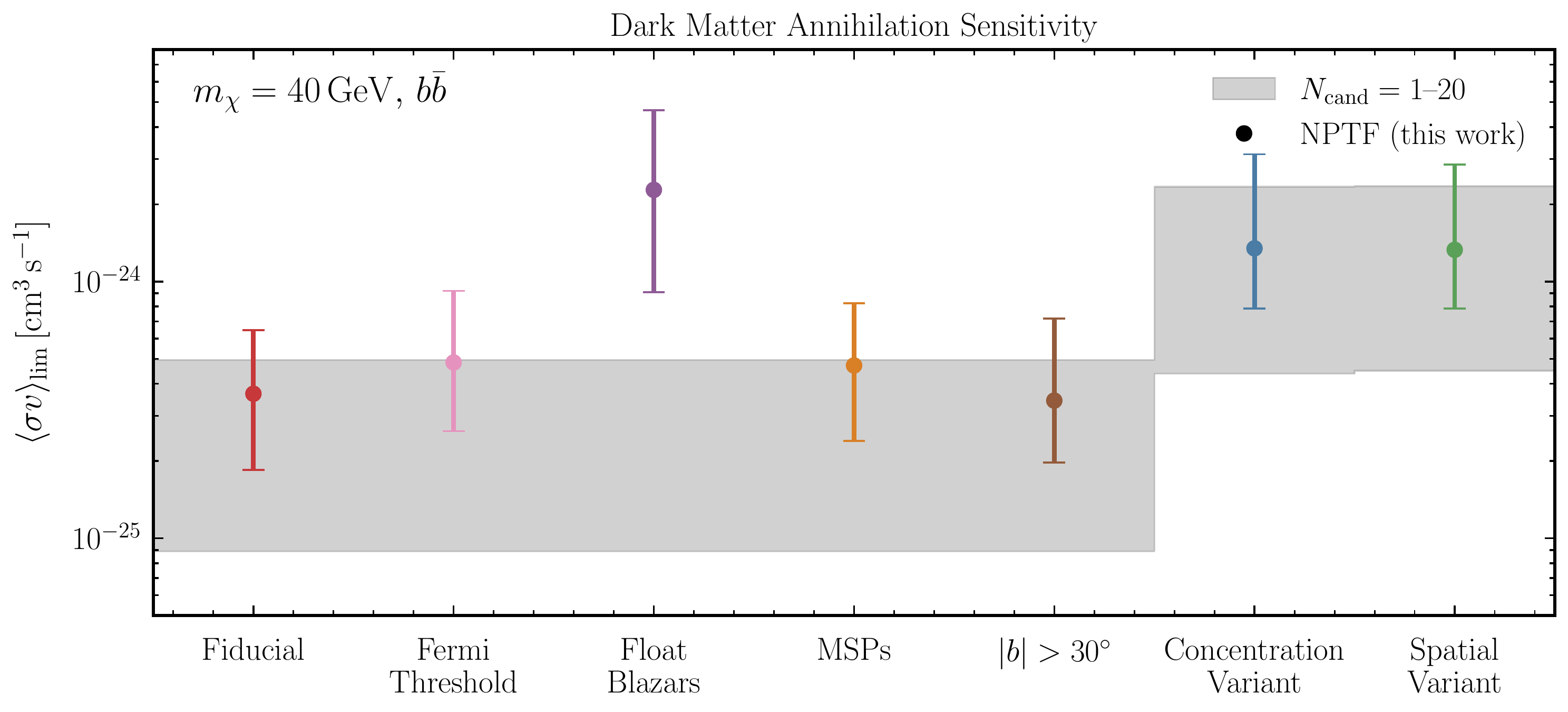}
\caption{Summary of how the projected limit on the annihilation cross section to $b\bar{b}$ changes for the different scenarios tested, for a fixed DM mass of $m_\chi = 40$\,GeV.  Each point denotes the middle 1-$\sigma$ containment of the 95\% confidence interval limit obtained over 200-300 Monte Carlo realizations.  The gray band indicates the corresponding limits from the traditional subhalo detection methods, assuming that the number of resolved subhalo candidates $N_\text{cand}$ falls in the range of  $1$ (lower edge of band) and $20$ (upper edge of band).  The ``Fiducial'' case (red point) corresponds to the limit recovered when using our fiducial subhalo and background models. The ``Fermi Threshold'' case (pink point) corresponds to when the fiducial scenario is run using the approximate 3FGL efficiency, effectively removing the lowest-flux sources.  The ``Float Blazars'' case (purple point) corresponds to being wholly model-independent in the treatment of blazars and fitting that background component with a broken power-law, assuming the Fiducial model.   The red and purple points bracket the most and least optimistic cases; in reality, we expect that the blazar model can be anchored by fixing some well-understood parameters of the model, while letting others float in the analysis.  The ``MSP'' case (orange point) corresponds to including pulsars in the analysis. The ``${\mid}b{\mid}>30^\circ$'' case (brown point) corresponds to restricting our ROI to higher-latitude regions. The Spatial and Concentration Variant (blue and green points, respectively) correspond to the variations on our Fiducial subhalo model, as described in Tab.~\ref{tab:models}, with the blazar model held fixed.
}
\label{fig:limit100}
\end{figure*}
\section{Results}
\label{sec:results}

 For a given subhalo and background model, the analysis procedure described in Sec.~\ref{sec:methodology} can be run on background-only simulated maps (\emph{i.e.}, containing no DM signal) in order to assess the sensitivity of our method and benchmark it against established techniques, such as searches for individual subhalo DM annihilation candidates. Figure~\ref{fig:limit100} shows the results of such analyses for the subhalo models, background models, and analysis variations considered in this work.  We present results for a benchmark DM mass of $m_\chi = 40$\,GeV, assuming annihilation to $b\bar b$.  For each subhalo and background model, we repeat the limit-finding procedure over 200 Monte Carlo realizations of the sky map.  Each point in Fig.~\ref{fig:limit100} shows the median of all the 95\% confidence limits obtained over the separate Monte Carlo realizations; the 1-$\sigma$ (middle-68\%) spread in the recovered limits is indicated by the error bars.\footnote{To ensure the validity and consistency of our statistical method, we perform a series of signal injection and recovery tests.  The results are presented in Fig.~\ref{fig:siginj}. } The traditional limits obtained from assuming $1$--$20$ resolved subhalo candidates, calculated using the procedure described in Section~\ref{sec:reslimit}, are shown as gray bands. In the rest of this section, we will describe the results for each model in detail.
 
We begin by considering the sensitivity to DM annihilation when we use our Fiducial subhalo model and assume perfect knowledge of the blazar background, shown as the red data point in Fig.~\ref{fig:limit100}. 
Comparing the point to the gray bands, we see that, within this Fiducial scenario, our method provides a complementary approach to search for DM annihilation in subhalos when there is some prior knowledge about the properties of astrophysical PSs.

 The next four points in Fig.~\ref{fig:limit100} show the results for variations in our background model and analysis configuration. First, in order to better understand the degree to which the unresolved sources affect the sensitivity, we repeat the analysis using the approximate 3FGL efficiency to exclude all detectable sources (astrophysical as well as subhalo). This contrasts with our fiducial scenario, where we have included sources that are unresolved, as well as those that are resolved but unassociated (which we mock-up by including all sources that contribute $\lesssim90$ photons). Including only unresolved sources results in only a factor of $\sim 2$ reduction in sensitivity compared to the fiducial setting, as seen from the pink data point in Fig.~\ref{fig:limit100}. This indicates that the population of unresolved sources contributes significantly to the achievable sensitivity.

\begin{figure}[t]
\centering
\includegraphics[width=0.45\textwidth]{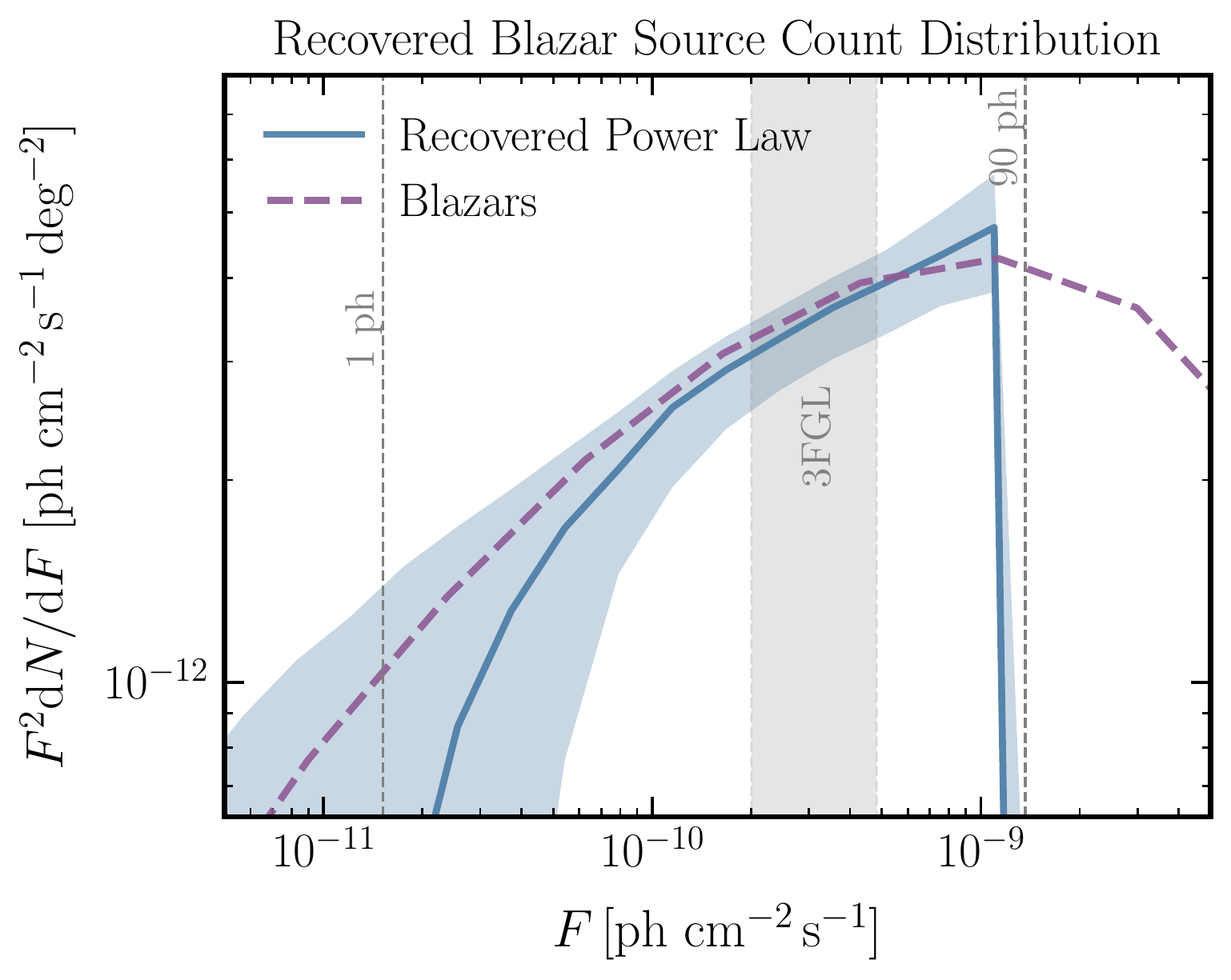}
\caption{Inferred triply-broken power-law blazar source-count distribution, showing median (blue line) and middle 1-$\sigma$ containment (blue band) over 300 realizations, for $\langle \sigma v \rangle = 10^{-24}\,{\rm cm}^3\,{\rm s}^{-1}$ and a DM mass $m_{\chi}=40$\,GeV. The true underlying blazar source-count distribution is shown as the dashed purple line. The recovered blazar source-count distribution compares favorably to the true underlying blazar source-count distribution. The grey band shows the approximate detection threshold of the 3FGL PS catalog obtained using the \emph{Fermi} source detection efficiency. The 1-photon line is the approximate threshold to which the NPTF is sensitive; the 90-photon line the approximate threshold above which all sources are assumed to be associated.}
\label{fig:blazar_scd}
\end{figure}

We can also assess how the limit is impacted when the blazar model is not fixed and is instead described by a triply-broken power law whose parameters are fit in the analysis.  The results, which assume the Fiducial subhalo model, are indicated by the purple point in Fig.~\ref{fig:limit100}.   In this case, we have increased the number of trials from $200$ to $300$ to ensure convergence of our results. As anticipated, weaker projected limits are obtained when we remain agnostic to the background PS population.  As we will discuss in the following section, this is suggestive of the fact that our limit is driven primarily by the differences in the blazar and subhalo source counts, as opposed to the differences in their spatial templates. Hence, once we remove prior knowledge about the blazar source-count distribution, our sensitivity is noticeably reduced. As indicated in Fig.~\ref{fig:blazar_scd}, in the presence of an injected DM signal with $\langle \sigma v \rangle = 10^{-24}$\,cm$^3$\,s$^{-1}$, the distribution of recovered source counts for the background PS model (blue band) is faithful to the true underlying blazar source-count distribution (purple line).

We note that while we have considered cases corresponding to \emph{(i)} assuming a particular form of the blazar contribution to the IGRB and \emph{(ii)} fitting a broken power law-parameterized source-count distribution to an unknown astrophysical background, in practice it may be possible and desirable to model the astrophysical background contribution in a physically-motivated manner. For example, Ref.~\cite{Manconi:2019ynl} measured the blazar source-count distribution by first assuming observationally motivated models for, \emph{e.g.}, the blazar luminosity function and spectral energy distribution and then fitting for \emph{only} the parameters of those models. Such a technique would allow us to incorporate prior information into the astrophysical source-count distribution, and would result in intermediate sensitivity to the scenarios considered here. 

Next, we consider the impact of including an unmodeled Galactic pulsar population in our sky map.  Although the overall flux of Galactic pulsars is expected to be subdominant to that of extragalactic blazars, they may be spatially more correlated with subhalos and introduce degeneracies in the photon-count analysis. We test this by adding a population of disk-correlated millisecond pulsars, modeled using the best-fit luminosity function and spatial profile obtained in Ref.~\cite{Bartels:2018xom}.  The MSP source-count distribution in the 2--20~GeV energy range is shown in the left panel of Fig.~\ref{fig:SCD_all}.  The MSP distribution is suppressed relative to the blazar distribution, and peaked several decades in flux below the 3FGL resolution threshold.  Because the MSPs do not contribute many high-flux sources in the high-latitude analysis ROI and energy-range of interest here, in addition to being spatially more concentrated at lower latitudes, we expect their impact on the overall limit to be relatively small. Indeed, we find that the resulting limit when including the unmodeled pulsar population for the $m_\chi= 40$\,GeV example, shown as the orange point in Fig.~\ref{fig:limit100}, is weakened by only a factor of $\sim1.5$ compared to the fiducial case. Note that it is likely that this weakening would be reduced if we used the ${\mid}b{\mid}>30^\circ$ mask described in the next variation and/or explicitly modeled the unresolved Galactic PS contribution.

One of the largest uncertainties in our analysis is our model of the diffuse gamma-ray emission. In an analysis on real \Fermi-LAT data, any mismodeling could induce spurious signals or background over-subtraction, especially near the edges of our analysis mask ($b\sim20^\circ$) where the diffuse emission is largest and most uncertain. We are also subject to the assumption that any emission from the Milky Way's smooth halo is negligible, and at the edges of our fiducial analysis mask, this emission can be comparable to the astrophysical Poissonian emission. Thus, we repeat our fiducial analysis with a latitude cut of ${\mid}b{\mid}>30^\circ$, removing a larger region around the Galactic plane and significantly reducing the impact of an uncertain diffuse emission and emission from an annihilating smooth Milky Way halo. The results are shown in Fig.~\ref{fig:limit100} as the brown data point, and do not change signifcantly compared to results with our fiducial ROI mask of ${\mid}b{\mid}>20^\circ$.

The variations on our background models demonstrate that our techniques are relatively robust to many of the dominant astrophysical uncertainties. We next consider systematic uncertainties in modeling the Galactic subhalo population. As summarized in Tab.~\ref{tab:models} and described in detail in Sec.~\ref{sec:shmodels}, we consider two variants of the Fiducial subhalo model.  The first (Concentration Variant) utilizes an alternative concentration model~\cite{Sanchez-Conde:2013yxa} that is distance-independent.  This model ignores the fact that subhalos closer to the Galactic Center are likely to be more concentrated due to tidal effects, as accounted for in the distance-dependent concentration model from Ref.~\cite{Moline:2016pbm} that is used in the Fiducial model.  The second variation (Spatial Variant) uses the subhalo number density distribution from the Phat-ELVIS simulations~\cite{Kelley:2018pdy}, which models the effect of baryons and hence includes enhanced subhalo disruption, especially close to the Galactic plane.

The result of either of these variations is a depressed population of brighter subhalos closer to the Solar radius and, consequently, reduced sensitivity reach compared to the fiducial case. Figure~\ref{fig:limit100} compares the annihilation cross section limit for both of these (blue data point for the Concentration Variant, green data point for the Spatial Variant) scenarios, with all other particulars of the modeling being the same as the fiducial case (red point).  We see that the projected limit is weakened by a factor of $\sim4$ in these cases, compared to the Fiducial model. However, the projected limits for the individual subhalo search method are also correspondingly weakened.  In each case, we find that our method can provide a complementary probe of DM annihilation from a subhalo population, especially when a relatively large number ($\gtrsim 10$) of unassociated candidate sources are present. This result makes sense, given that at the higher cross sections where $\gtrsim 10$ subhalos are resolvable, we expect the unresolved subhalo component to correspondingly contribute a larger, and hence more easily detectable, signal.

\begin{figure*}[t]
\centering
\includegraphics[width=0.47\textwidth]{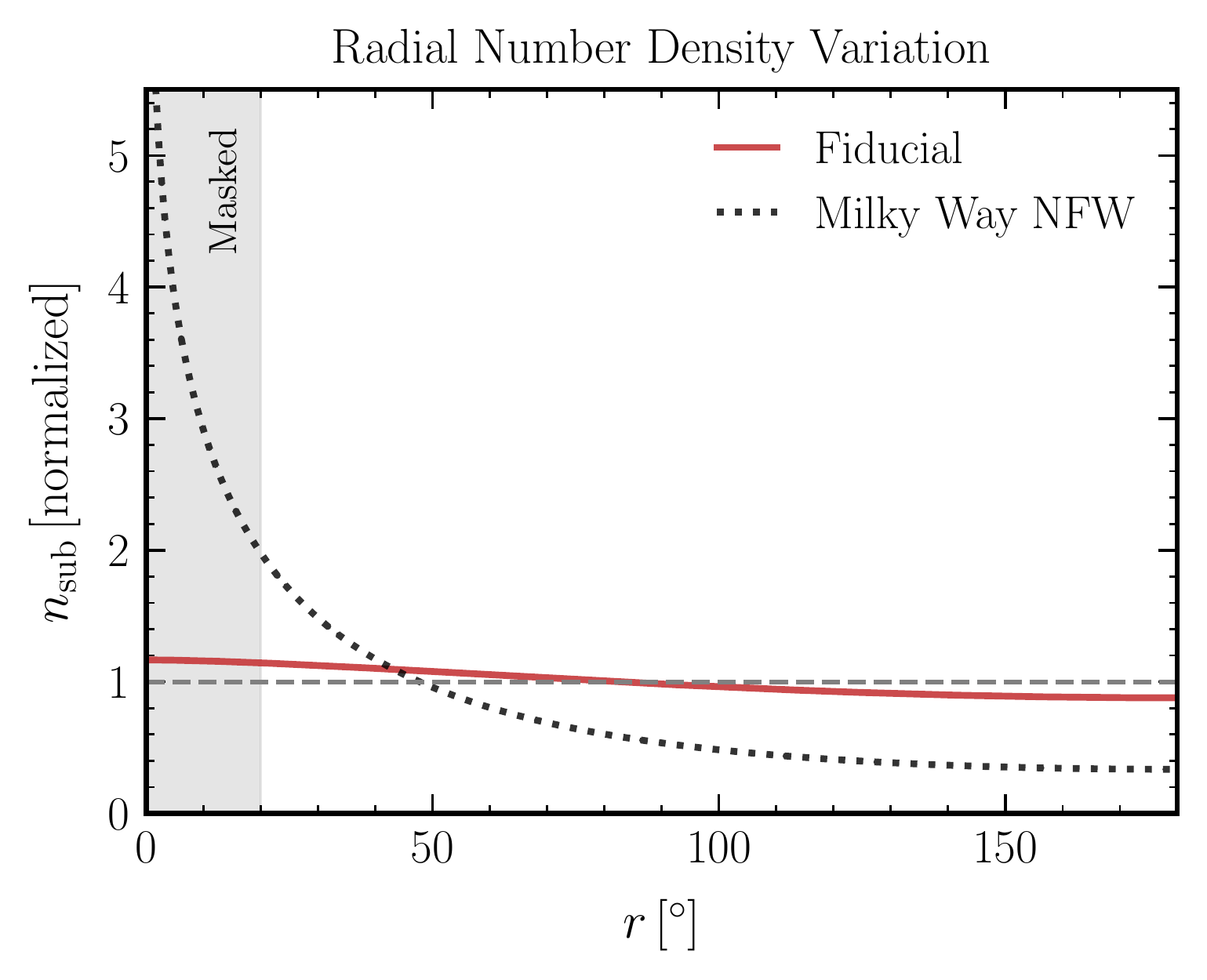}
\includegraphics[width=0.47\textwidth]{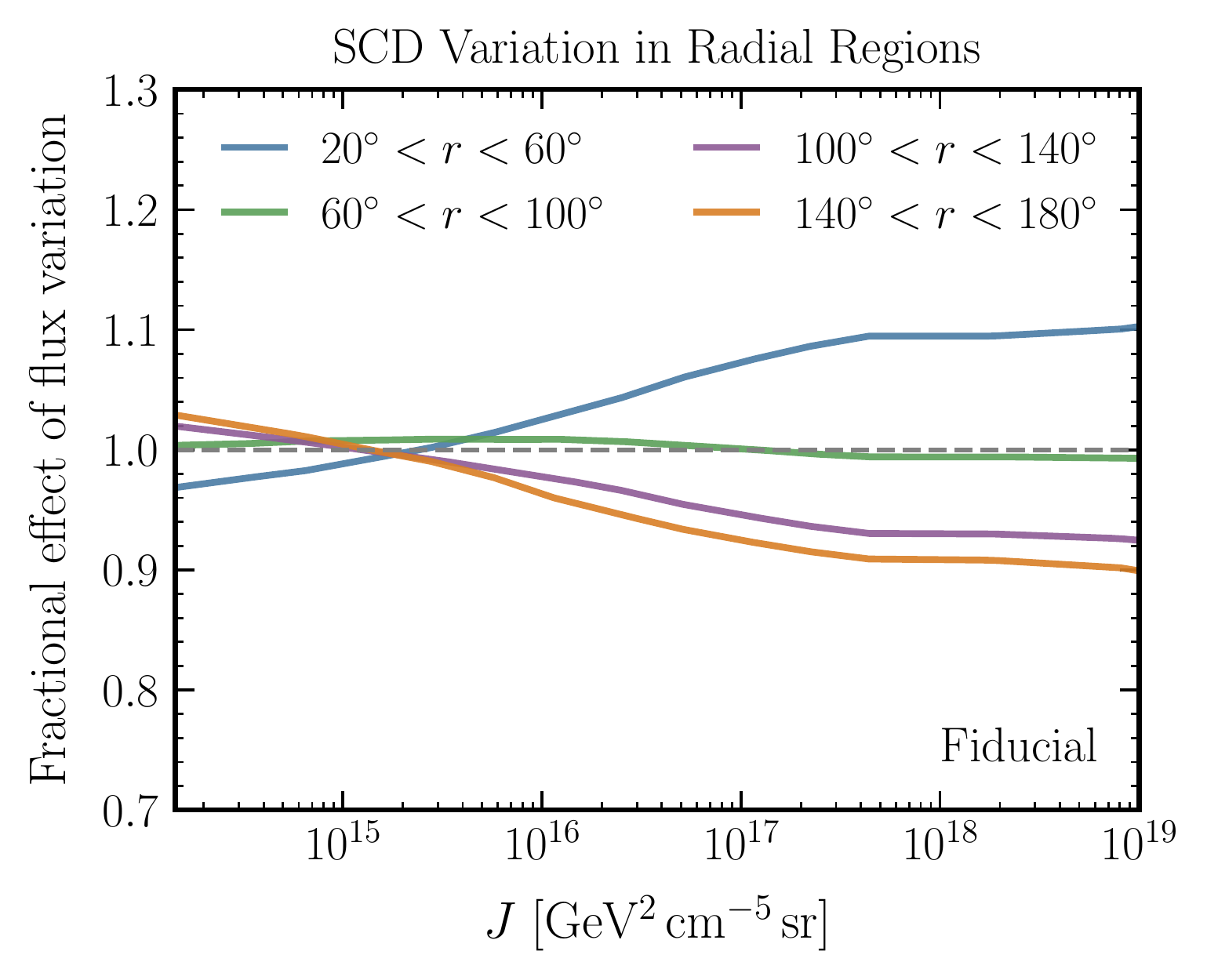}
\caption{\emph{(Left)} Variation in the number density of subhalos as a function of angular distance from the Galactic Center, shown for our Fiducial model (red line) and an infall subhalo population spatially following the smooth Milky Way halo (dotted black line). This shows the effect captured by the non-Poissonian spatial template. \emph{(Right)} Ratio of the differential source-count distributions (SCDs) as a function of subhalo $J$-factor in four radial regions to the sky-averaged source-count distribution, shown for the Fiducial subhalo model (solid lines). Differences in the distribution of subhalo properties---in this case, the concentration and distance to the subhalo---along different lines of sight introduce  variations in the source count in the different regions.}
\label{fig:spatial_flux_variation}
\end{figure*}

\section{Distinguishing features of a subhalo population}
\label{sec:fluxreg}

We have so far treated the non-trivial spatial distribution of Galactic subhalos across the sky (modeled through the non-Poissonian template) and the distribution of their ROI-averaged expected flux contribution due to DM annihilation (modeled through the source-count distribution) as handles to distinguish subhalos from an astrophysical PS population. In this section, we discuss how much discriminatory power each of these lends to the analysis, and also explore an additional discriminator that can be used to identify an annihilating subhalo population: the expected variation of the subhalo source-count distribution along different lines-of-sight.

Figure~\ref{fig:spatial_flux_variation} (left) shows the normalized angular number density of subhalos as a function of radial angle from the Galactic Center, for our Fiducial subhalo configuration (red line) and an infall population that spatially follows the smooth Milky Way halo (black line).  As already discussed in relation to Fig.~\ref{fig:maps} (left), the spatial distribution of subhalos is very close to isotropic, deviating from a spatially isotropic population (dashed gray line) by $\mathcal{O}(10\%)$. This is indicative of the fact that the distinct spatial angular number density distribution of Galactic subhalos, as captured by the non-Poissonian template, is not a significant distinguishing feature in the face of an isotropic, extragalactic population of astrophysical PSs. This hypothesis is supported by the weakening of our limit when we float the blazar source-count distribution, as was described in Sec.~\ref{sec:results}. In fact, we have verified that using an isotropic distribution to characterize the subhalo population does not change our fiducial results. The primary signal discriminator in our analysis is therefore the overall flux and its distribution, for both resolvable and unresolved subhalos, as captured by the source-count distribution.

In addition to these features, the source-count distribution of subhalos is also expected to vary along different lines of sight in a non-trivial manner due to variations in, \emph{e.g.}, the subhalo line-of-sight distances and virial concentrations along different directions. To make this concrete, consider two opposing lines of sight---one towards the Galactic Center and the other away from it. Due to the larger expected number density of subhalos towards the inner regions of the Milky Way, we expect a larger fraction of sources with small line-of-sight distances when looking towards the Galactic Center rather than away from it (normalized to the same number of sources in each direction). Additionally, tidal forces are responsible for subhalos close to the center of the Galaxy being on average more concentrated, leading to a larger fraction of subhalos with higher concentrations towards the Galactic Center. Both of these effects result in a larger fraction of subhalos with higher $J$-factors, and a source-count distribution correspondingly skewed towards having  brighter sources in radial directions closer to the Galactic Center.

To quantify this effect, Fig.~\ref{fig:spatial_flux_variation} (right) shows the fractional variation in the source-count distribution in four radial regions as a function of subhalo $J$-factor. It can be seen that the different subhalo property distributions in different radial directions result in an additional $\mathcal O(10\%)$ variation in the source-count distribution of subhalos across the sky on top of the variation in subhalo number density, itself an $\mathcal O(10\%)$ effect. We note that the vast majority of this variation is sourced by variations in the concentration distribution of subhalos in different radial directions rather than their varying line-of-sight distance distributions.

Flux variation can be captured in the non-Poissonian analysis framework by considering separate source-count distributions in different radial rings and summing the independent log-likelihoods together. For our fiducial set-up, using 20 radial rings within our ROI, we find that considering flux variation makes no appreciable difference to the overall results. This further underscores the fact that the nearly-isotropic spatial distribution of subhalos is not a significant discriminator in this analysis. However, we note that in specific applications and non-standard scenarios---such as when considering a more centrally-concentrated substructure population---flux variation across the ROI could be an important effect.

\section{Conclusions}
\label{sec:conclusions}

In this paper, we have introduced a method that leverages photon-count statistics to search for a DM annihilation signal from a population of Galactic subhalos. We have shown that the subhalo population can collectively leave a unique, detectable imprint on the distribution of photons in gamma-ray data that can be searched for using the Non-Poissonian Template Fitting (NPTF) framework, even when subhalos cannot be individually resolved as high-significance PSs. We demonstrated the feasibility of our method using simulated \emph{Fermi}-LAT data, showing in particular that the annihilation signal can be searched for even in the presence of large backgrounds expected from unresolved PSs of extragalactic and Galactic origin such as blazars and pulsars, respectively. While we have tested our method with particular choices of subhalo models and within a frequentist analysis framework, modifications for arbitrary assumptions about the subhalo population and using other means of statistical inference are easily admitted. In particular, a Bayesian analysis framework could allow for the inclusion of prior information about the background PS population in a straightforward manner. 

While methods using gamma-ray 1-point statistics over large regions of the sky can be susceptible to the effects of Galactic diffuse mismodeling, foreground contamination is expected to be less of an issue at the higher latitudes considered here. In order to quantitatively assess the impact of diffuse mismodeling, we repeated our fiducial study using the \texttt{p6v11}\footnote{\url{https://fermi.gsfc.nasa.gov/ssc/data/access/lat/ring_for_FSSC_final4.pdf}\url{}} diffuse model in the template analysis while still simulating the gamma-ray map using diffuse \texttt{Model A}. This test represents a particularly serious case of diffuse mismodeling at high latitudes, since the two models were produced and optimized under significantly different assumptions (see Refs.~\cite{Calore:2014xka,Ackermann:2014usa} for further details). We find that this results in a weakening of the overall sensitivity by a factor of $\sim 2$, a surprisingly small reduction considering the extreme (and likely unrealistic)  mismodeling in this test.  However, these results do highlight that a robust analysis on data would have to consider variations on the diffuse foreground model and/or use data-driven methods for mitigating mismodeling effects~\cite{Buschmann:2020adf,Chang:2018bpt,Storm:2017arh,Selig:2014qqa}. Additionally, while the model of blazar-like emission that we consider in our proof-of-principle study is meant to be representative, departures from this model can affect the results presented here.

Our method is complementary to established techniques, including those relying on detecting individually resolved DM annihilation subhalo candidates. By definition, our technique is less impacted by the presence of astrophysical but not yet associated bright candidate sources that can significantly weaken the constraints obtained using established searches. Moreover, it can take into account both the putative resolved and unresolved subhalo populations. We have shown that, with our choice of models for the subhalo and background astrophysical populations, our method can lead to stronger constraints on annihilation properties when $\gtrsim \mathcal O(10)$ candidate subhalos are detected.  The success of the method relies on the discriminatory power of the DM and background source-count distributions.  If a large number of candidate subhalos without associations are found in recent (\emph{e.g.,} the 4FGL~\cite{Fermi-LAT:2019yla}) and future PS catalogs, methods including sub-threshold emission could provide decisively better constraints on annihilation properties.  While the analysis presented in this paper focused on \emph{Fermi}-LAT searches, it can also be applied to other observations such as those from the upcoming Cherenkov Telescope Array (CTA)~\cite{2011ExA....32..193A,Hutten:2016jko,Hutten:2018wop} and proposed sub-GeV gamma-ray missions~\cite{Chou:2017wrw,DeAngelis:2017gra,Moiseev:2015lva,McEnery:2019tcm}.

Our proposed method is sensitive to cross sections in the $\langle \sigma v \rangle \sim 10^{-24}$--$10^{-25}$\,cm$^3$\,s$^{-1}$ range for the benchmark parameter point of $m_\chi=40$\,GeV and $b\overline{b}$ annihilation, consistent with and complementary to the sensitivity of established \emph{Fermi}-LAT subhalo searches.  These are, in general, weaker than the cross section limits from annihilation searches in dwarf galaxies~\cite{Fermi-LAT:2016uux}.  However, given the systematic uncertainties associated with DM annihilation limits in any given target, it is important to optimize sensitivities over a collection of independent search targets.  In this sense, annihilation searches in subhalos---as well as other targets such as \emph{e.g.}, galaxy groups~\cite{Lisanti:2017qlb,Lisanti:2017qoz,Ackermann:2010rg} and the Milky Way halo itself~\cite{Chang:2018bpt,Huang:2015rlu,Ackermann:2012rg}---should be pursued in tandem with dwarf studies.

Several extensions to the current framework are possible. In particular, our method as presented neglects spectral information, instead relying exclusively on the analysis of photon counts in a single energy bin. Extensions of the Non-Poissonian Template Fitting framework to an energy-binned analysis would allow one to leverage the unique spectral properties of a DM annihilation signal due to a Galactic subhalo population. Additionally, spatial extension of subhalos beyond the detector PSF~\cite{Coronado-Blazquez:2019pny} can be used as an additional handle and implemented within the NPTF framework as an ``effective PSF'' for the subhalo population (although we note that a larger effective PSF could exacerbate issues associated with diffuse mismodeling~\cite{Chang:2019ars}).  These extensions could lead to even better prospects for sub-threshold subhalo searches than those presented in this paper and we leave their study to future work. 
\vspace{0.1in}

\section{Acknowledgements}
\label{sec:acknowledgements}

We thank F.~Calore, C.~Combet, K.~Cranmer, M.~H\"{u}tten, and B.~Safdi for useful conversations. We also acknowledge L.~Iliesiu and Y.~Kahn for collaboration in the initial stages of this work.  LJC is supported by a Paul \& Daisy Soros Fellowship and an NSF Graduate Research Fellowship under Grant Number DGE-1656466. ML is supported by the DOE under Award Number DESC0007968 and the Cottrell Scholar Program through the Research Corporation for Science Advancement. SM is supported by the NSF CAREER grant PHY-1554858, NSF grants PHY-1620727 and PHY-1915409, and the Simons Foundation.  This work was performed in part at the Aspen Center for Physics, which is supported by National Science Foundation grant PHY-1607611.  The work presented in this paper was performed on computational resources managed and supported by Princeton Research Computing, a consortium of groups including the Princeton Institute for Computational Science and Engineering (PICSciE) and the Office of Information Technology's High Performance Computing Center and Visualization Laboratory at Princeton University. This work made use of the NYU IT High Performance Computing resources, services, and staff expertise. This research made use of the \texttt{CLUMPY}~\cite{Hutten:2018aix,Bonnivard:2015pia,Charbonnier:2012gf}, 
\texttt{corner}~\cite{corner}, \texttt{healpy}~\cite{2005ApJ...622..759G,Zonca2019}, \texttt{IPython}~\cite{PER-GRA:2007}, \texttt{Jupyter}~\cite{Kluyver2016JupyterN}, \texttt{matplotlib}~\cite{Hunter:2007}, \texttt{NumPy}~\cite{numpy:2011}, \texttt{SciPy}~\cite{Jones:2001ab}, and \texttt{tqdm}~\cite{da2019tqdm}  software packages.

\bibliographystyle{apsrev4-1}
\bibliography{subhalos_nptf}

\onecolumngrid

\newpage

\appendix
\section{Supplementary Figure}
\label{app:appendix}

\setcounter{equation}{0}
\setcounter{figure}{0}
\setcounter{table}{0}
\setcounter{section}{0}
\makeatletter
\renewcommand{\theequation}{S\arabic{equation}}
\renewcommand{\thefigure}{S\arabic{figure}}
\renewcommand{\thetable}{S\arabic{table}}

\begin{figure}[h]
\centering 
\includegraphics[width=0.45\textwidth]{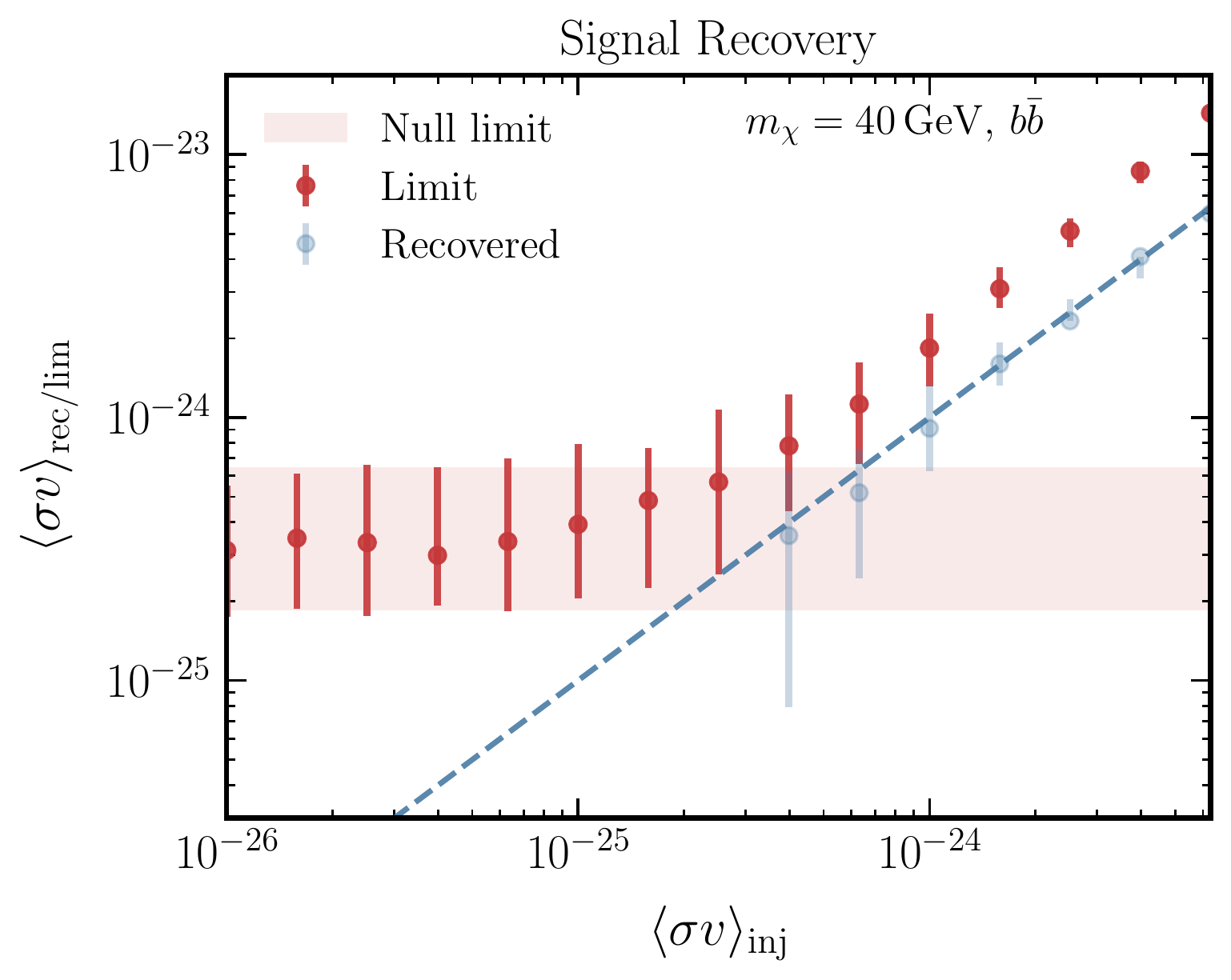}
\caption{Middle 1-$\sigma$ containment of the recovered cross section (blue points) and 95\% confidence interval limit (red points) for 200 Monte Carlo realizations as a function of injected DM cross section for our fiducial set-up.  The injected cross section is correctly recovered over the entire applicable cross section range, and the limit never excludes the injected signal. Recovered cross sections are not shown when the median significance of detection corresponds to a test statistic $\mathrm{TS} < 1$.  The one-to-one line is shown in dashed blue.  These tests confirm that the signal recovery and limit-setting procedure is working as desired.}
\label{fig:siginj}
\end{figure}

\end{document}